\documentclass[iop]{emulateapj}
\usepackage{graphicx}
\usepackage{epsfig}
\usepackage{natbib}
\begin{document}

\title{A Study of Carbon Features in Type Ia Supernova Spectra}

\author{Jerod T.\ Parrent\altaffilmark{1}, R.\ C.\ Thomas\altaffilmark{2},  
Robert A.\ Fesen\altaffilmark{1}, G.\ H.\ Marion\altaffilmark{3,4}, Peter Challis\altaffilmark{3}, Peter M.\ Garnavich\altaffilmark{5}, Dan Milisavljevic\altaffilmark{1}, J\`{o}zsef Vink\`{o}\altaffilmark{6}, J.\ Craig Wheeler\altaffilmark{4}}
\altaffiltext{1}{6127 Wilder Lab, Department of Physics \& Astronomy, Dartmouth
                 College, Hanover, NH 03755, USA.}
\altaffiltext{2}{Physics Division, Lawrence Berkeley National Laboratory, 1 Cyclotron Road, Berkeley, CA 94720, USA.}
\altaffiltext{3}{Harvard-Smithsonian Center for Astrophysics, 60 Garden St., Cambridge, MA 02138, USA.}
\altaffiltext{4}{Astronomy Department, University of Texas at Austin, Austin, TX 78712, USA.}
\altaffiltext{5}{Physics Department, University of Notre Dame, Notre Dame, IN 46556, USA.}
\altaffiltext{6}{Department of Optics \& Quantum Electronics, University of Szeged, D\`{o}m t\`{e}r 9, Szeged H-6720, Hungary.}
\slugcomment{Submitted to the Astrophysical Journal 2010 December 1; accepted 2011 February 28}

\shorttitle{Carbon Signatures in SNe Ia}
\shortauthors{Parrent et al.}

\begin{abstract}

One of the major differences between various explosion scenarios of Type Ia
supernovae (SNe Ia) is the remaining amount of unburned (C+O) material and its velocity
distribution within the expanding ejecta. While oxygen absorption features are
not uncommon in the spectra of SNe Ia before maximum light, the presence of
strong carbon absorption has been reported only in a minority of objects,
typically during the pre-maximum phase. The reported low frequency of
carbon detections may be due to low signal-to-noise data, low abundance
of unburned material, line blending between \ion{C}{2} $\lambda$6580 and
\ion{Si}{2} $\lambda$6355, ejecta temperature differences, asymmetrical
distribution effects, or a combination of these. However, a survey of published
pre-maximum spectra reveals that more SNe Ia than previously thought may exhibit
\ion{C}{2} $\lambda$6580 absorption features and relics of line blending near
$\sim$6300 \AA. Here we present new SN Ia observations where
spectroscopic signatures of \ion{C}{2} $\lambda$6580 are detected, and
investigate the presence of \ion{C}{2} $\lambda$6580 in the optical spectra of
19 SNe Ia using the parameterized spectrum synthesis code, \texttt{SYNOW}. Most
of the objects in our sample that exhibit \ion{C}{2} $\lambda$6580 absorption
features are of the low-velocity gradient subtype. Our study indicates that the
morphology of carbon-rich regions is consistent with either a spherical
distribution or a hemispheric asymmetry, supporting the recent idea that SN Ia
diversity may be a result of off-center ignition coupled with observer
line-of-sight effects. 

\end{abstract}


\section{Introduction}

The typical pre-maximum SN Ia spectrum consists of overlapping P-Cygni profiles
of intermediate-mass elements (IMEs) and Fe-peak elements (IPEs) that indicate
expansion velocities on the order of 10$^{4}$ km s$^{-1}$ \citep{Flipper97}.
Photometric properties of the rise, peak, and decline of a SN Ia light curve
can be explained by assuming that a substantial amount of $^{56}$Ni,
synthesized in the explosion, powers the SN luminosity
\citep{Colgate69,Arnett82}.  Consequently, results of spectroscopic and
photometric studies have supported the idea that SNe Ia are the outcome of a
thermonuclear explosion of a C+O white dwarf in a binary system
\citep{HF60,Nomoto84,Elias85,Iben88,Nomoto03,Chen09,Howell10}. 

The explosion mechanisms that have been proposed differ by how the
thermonuclear flame is propagated through the star's interior; i.e., a sub-sonic
deflagration via thermal conductivity or a super-sonic detonation due to strong
shock burning. Pure detonation models appear unlikely since they conflict with
observations by producing too much $^{56}$Ni, not enough IMEs, and leaving very
little unburned material behind \citep{Arnett69,Branch95}. On the other hand,
while pure deflagration models may account for fainter SN Ia events, they are
energetically weak, leaving too much material unburned to represent the
majority of SNe Ia \citep{Travaglio04,Gamezo05,Kozma05}. Thus, a
deflagration that transitions into a detonation may be necessary in order to
attain the nucleosynthetic yields that are consistent with the observations
\citep{Khokhlov91,Hoflich95,Kasen09,Maeda10}. 

Additional facets of modeling (e.g., multidimensional considerations) will
affect the abundance tomography as well \citep{Woosley09}. This can make it
difficult to distinguish the dominant sources of SN Ia diversity. For certain,
however, one similarity between all explosion models is the existence of burned
(ash) and unburned material (fuel). By using spectroscopic signatures of C and
O at pre-maximum epochs, one can infer the amount and velocity structure of the
outer, unburned material. 

Some of the unburned material may be subject to a degree of downward mixing
toward the inner ejecta when burning becomes turbulent and enters the
distributed flame regime \citep{Pope87,Niemeyer98,Aspden10}. As of a result, it
is not certain whether regions of unburned material are strictly located in
the outer layer or mixed within the rest of the ejecta, nor is the mass range
of unburned material from these thermonuclear explosions known \citep{Baron03}.
Therefore, the rate of detection of unburned material and its phenomenological
details is of great importance for constraining how much material remains
unburned in the explosion models. 

The most prominent oxygen line in the optical spectra of SNe Ia is \ion{O}{1}
$\lambda$7774. Unfortunately, since oxygen is a product of carbon burning,
oxygen absorption lines are likely to be a biased reference for measuring the
location of unburned material. Thus one must look to carbon absorption features
as a tracer for unburned material. While it is not uncommon for oxygen to be
present in SN Ia spectra \citep{Branch06}, signatures of carbon have only been
sporadically reported in SN Ia spectra, most often during the pre-maximum
phase.

For the typical ejecta temperatures seen in SNe Ia ($\sim$10,000 K), the
dominant ionization state of carbon is \ion{C}{2} \citep{Tanaka08}. At this
temperature, the strongest optical line of \ion{C}{2} is that of $\lambda$6580
\citep{Hatano99b}. When seen, this line produces a blue-shifted absorption near
6300 \AA\ that sometimes blends with the neighboring emission component of the
\ion{Si}{2} $\lambda$6355 P-Cygni profile, thus making \ion{C}{2}
identifications problematic. There are well observed SNe Ia where
\ion{C}{2} $\lambda$6580 is clearly present
\citep{Patat96,Mazzali01,Garavini05,Hicken07,Thomas07,Yamanaka09b,Scalzo10}, however in
general, \ion{C}{2} absorption is often weak and/or blended and
therefore not a conspicuous SN Ia feature \citep{Salvo01,Branch03,Stanishev07}. 

A number of factors intrinsic to the explosion itself may contribute to the
strength of \ion{C}{2} $\lambda$6580 and whether or not carbon features show up
in SN Ia spectra. These include the asymmetrical distribution of carbon, the
extent of carbon burning, the temperature of the carbon-rich region, and the
possible formation of an envelope of unburned material though a merger of two
white dwarfs \citep{Thomas07,Tanaka08,Scalzo10}. Low signal-to-noise spectra
and effects of line formation may also obscure weak \ion{C}{2} $\lambda$6580
absorption features, thereby complicating the rate of detection further.

In an attempt to better understand spectroscopic carbon signatures in the
pre-maximum optical spectra of SNe Ia, we present a comparative study of 16 SNe
Ia where a \ion{C}{2} $\lambda$6580 absorption signature is evident, plus three
additional cases where it may be present. Then using the spectrum synthesis
code, \texttt{SYNOW}, we fit this sample of SN Ia spectra thereby mapping the
velocity distribution of carbon-rich regions that give rise to \ion{C}{2}
$\lambda$6580 absorption features. To ensure a consistent analysis, we
generated synthetic spectra for each of the time-series spectra in our sample.

The new and archival data used in this study are presented in \S2. We review SN
Ia subtype classifications in \S3, and in \S4 we discuss the spectrum fitting
methods implemented with the \texttt{SYNOW} model, followed by our results in
\S5. We discuss our results in \S6 in the context of recent findings on the
diversity of SNe Ia and conclude in \S7 with our results regarding of the
nature of \ion{C}{2} absorption features. 

\section{Data}

Low signal-to-noise (S/N) SN Ia spectra make investigations of any suspected
\ion{C}{2} absorption features difficult. Therefore, with the possibility of
carbon features being both weak and blended with \ion{Si}{2} $\lambda$6355,
well observed multi-epoch confirmations of any suspected \ion{C}{2}
$\lambda$6580 signatures were preferred for analysis.

In constructing our data sample, we have included only SNe Ia where at least
two consecutive spectra showed the presence of a $\sim$6300 \AA\ absorption
signature, indicating a likely \ion{C}{2} $\lambda$6580 absorption feature.  In
addition, if the \ion{C}{2} feature was not seen before maximum light, then we
did not include those SNe in our sample since this region of the spectrum
becomes contaminated by neighboring \ion{Fe}{2} lines within a week after
maximum light \citep{Branch08}. The two exceptions are the single epoch spectra
of PTF10icb and the 2002cx-like, SN 2008ha, where the \ion{C}{2}
$\lambda$6580 signature is evident.

\subsection{New Observations}

\begin{figure}
\begin{center}
\includegraphics[width=\linewidth]{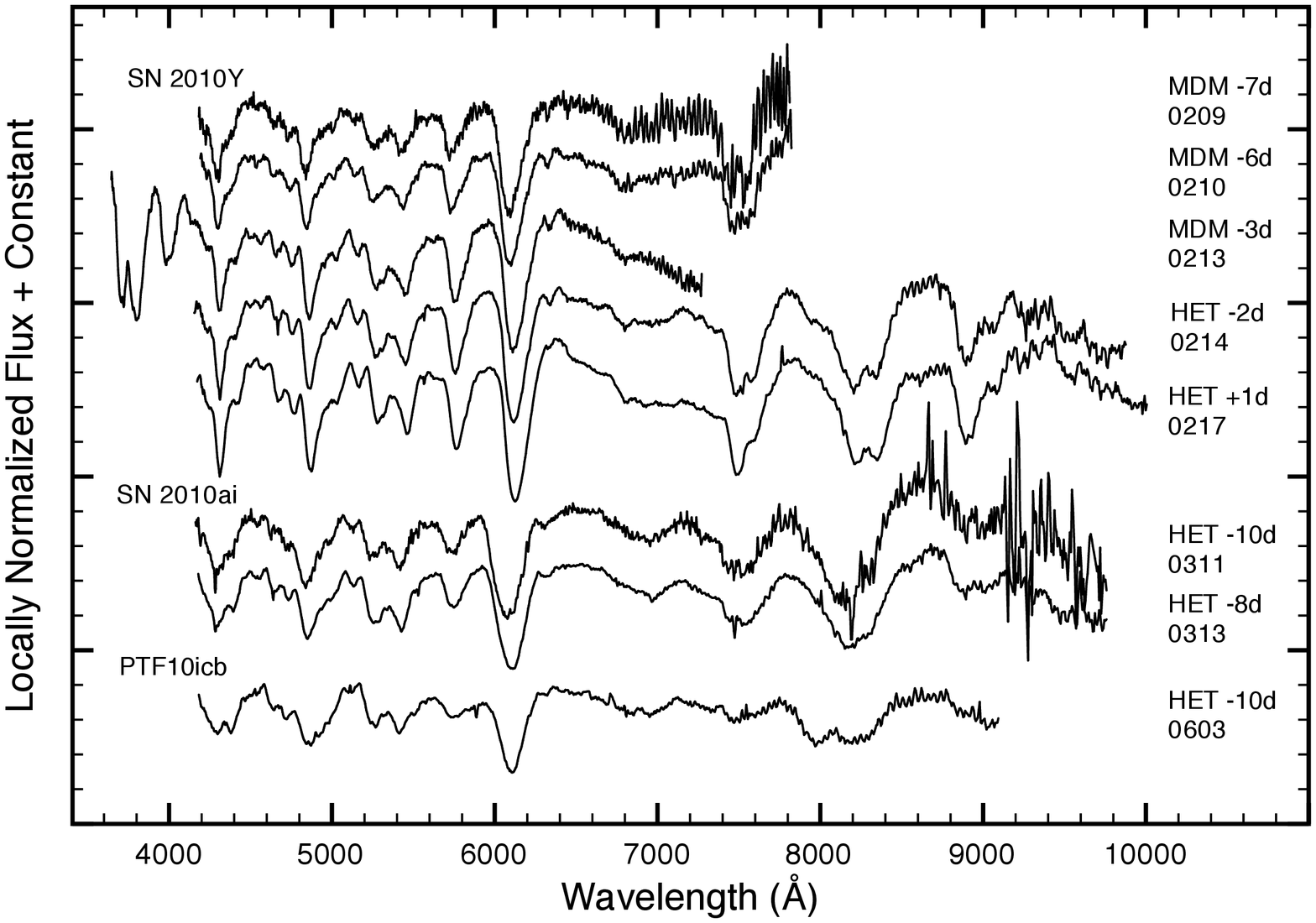}
\end{center}
\caption{Pre-maximum spectra of SN 2010Y, 2010ai, and PTF10icb. Data are from HET and the 2.4 m Hiltner at MDM, have been scaled appropriately, and are in the rest frame of their respective host galaxy. Absorption features that are consistent with \ion{C}{2} $\lambda$6580 blue-shifted to typical expansion velocities of SNe Ia are seen at $\sim$6300 \AA. See Table 2 for observational details.}
\label{fig:data}
\end{figure}

Pre-maximum optical spectra of SN 2010Y, 2010ai, and PTF10icb show absorptions
likely due to \ion{C}{2} $\lambda$6580 and thus we included these in our
sample. These data were reduced using standard IRAF procedures, were corrected
for host galaxy redshift and are presented in Figure~\ref{fig:data} and Table
1. We used the spectrum-comparison tool, SNID \citep{BT07}, to estimate the age
of each spectrum relative to maximum light. 

Three spectra for SN 2010Y covering days $-$7, $-$6, and $-$3 with respect to
maximum light were obtained using a Boller \& Chivens CCD spectrograph (CCDS)
on the 2.4 m Hiltner telescope at the MDM Observatory on Kitt Peak, Arizona.
These spectra show the emergence of a relatively weak yet persistent 6330 \AA\
absorption feature. Two additional spectra were taken on day $-$2 and $+$1 and
were acquired with the 9.2 m Hobby-Eberly Telescope (HET) \citep{Ramsey98} at
the McDonald Observatory using the Marcario Low-Resolution Spectrograph (LRS,
\citealt{Hill98}). By day $-$2, the 6330 \AA\ feature peaked in intensity and
began to fade by day $+$1. 

Low resolution optical spectra of SN 2010ai were also obtained with HET.
Spectra were taken well before maximum light on days $-$10 and $-$8. These
spectra show signs of a flux depression in the emission component of the
\ion{Si}{2} $\lambda$6355 P-Cygni profile, suggesting the presence of
\ion{C}{2} $\lambda$6580 absorption. 

Additional data comes from the discovery of PTF10icb by The Palomar Transient
Factory \citep{Nugent10}. A low-resolution follow-up spectrum was obtained with
the LRS on HET on June 3. The spectrum-comparison tool, SNID, identifies the
spectrum of PTF10icb as that of a normal SN Ia near day $-$10. Similar
to SN 2010Y and 2010ai, the spectrum of PTF10icb also exhibited a 6330 \AA\
absorption feature. 

\subsection{Archival Data}

In Table 2, we list these three recent objects along side 65 SNe Ia found in
the literature with pre-maximum or near maximum light spectra. These 68 SNe
have been organized by the subtype scheme of \cite{Benetti05} and are labeled
by the classification subtypes of \cite{Branch06} (see \S3). Since the
progenitor channel and the origin of \ion{C}{2} $\lambda$6580 absorption
features may differ for Super-Chandra SNe Ia, we separate SN 2003fg, 2006gz,
2007if and 2009dc as possible Super-Chandra candidates. We have also grouped SN
2000cx, 2002bj, 2002cx, 2007qd and 2008ha as miscellaneous objects since their
classifications as SN Ia-like events are still under debate
\citep{Valenti09,McClelland10,Poznanski10}. 

Optical spectra for all but three SNe Ia in our sample were obtained from the
Supernova Spectrum Archive
(SuSpect\footnote[1]{http://nhn.ou.edu/$\sim$suspect}) and sources therein. All
spectra have been corrected for host galaxy redshift and normalized according
to the formula given by \cite{Jeffery07} in order to remove the underlying
continuum since in this study we are only concerned with the position of the
\ion{C}{2} $\lambda$6580 absorption minimum and not absolute flux. The
inclusion of SN 2008ha is for comparison purposes only and
not meant to contribute to the discussion of carbon-rich regions of normal SNe
Ia. 

\begin{figure}
\centering
\includegraphics[width=13.8cm]{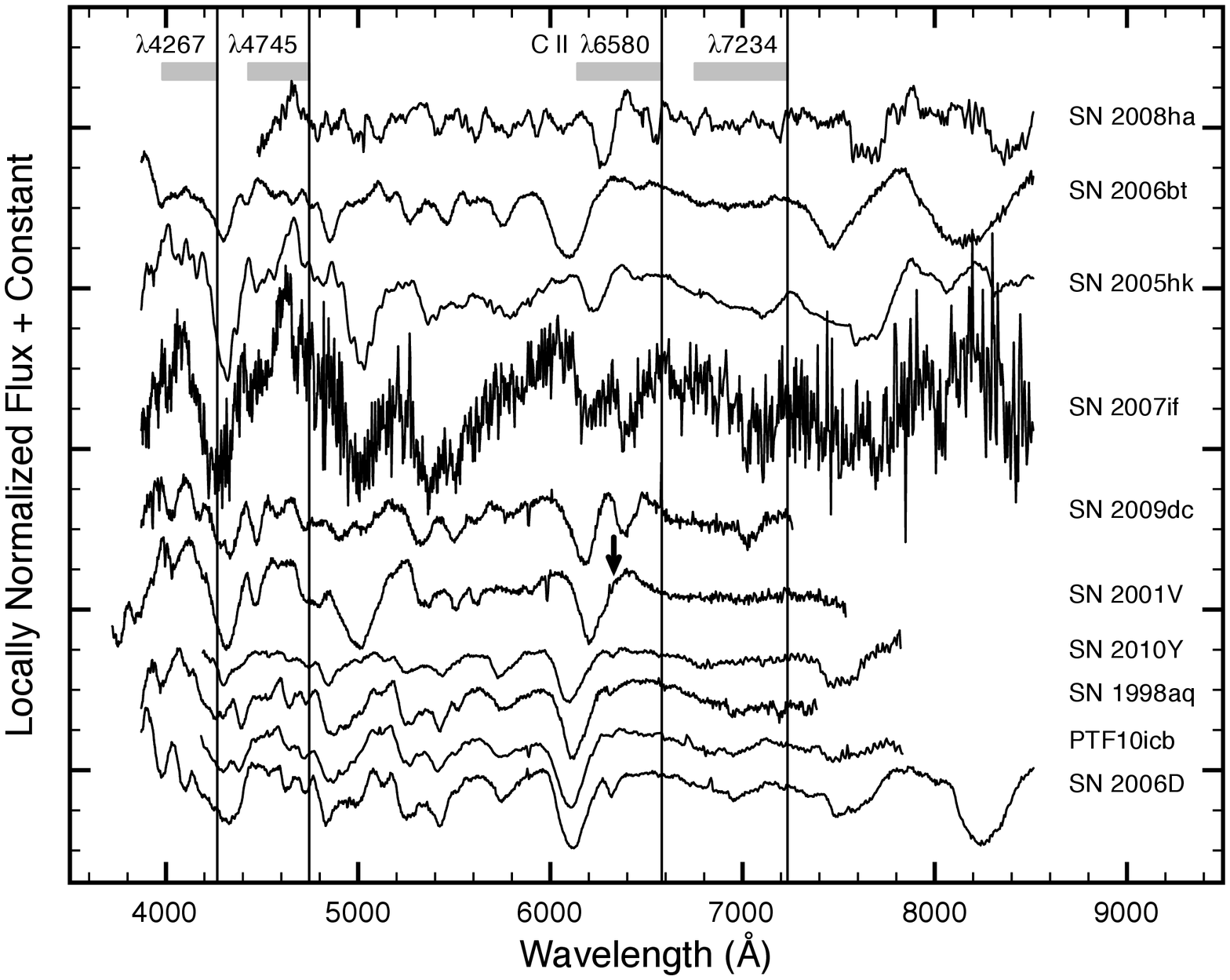}
\centering
\includegraphics[width=14.0cm]{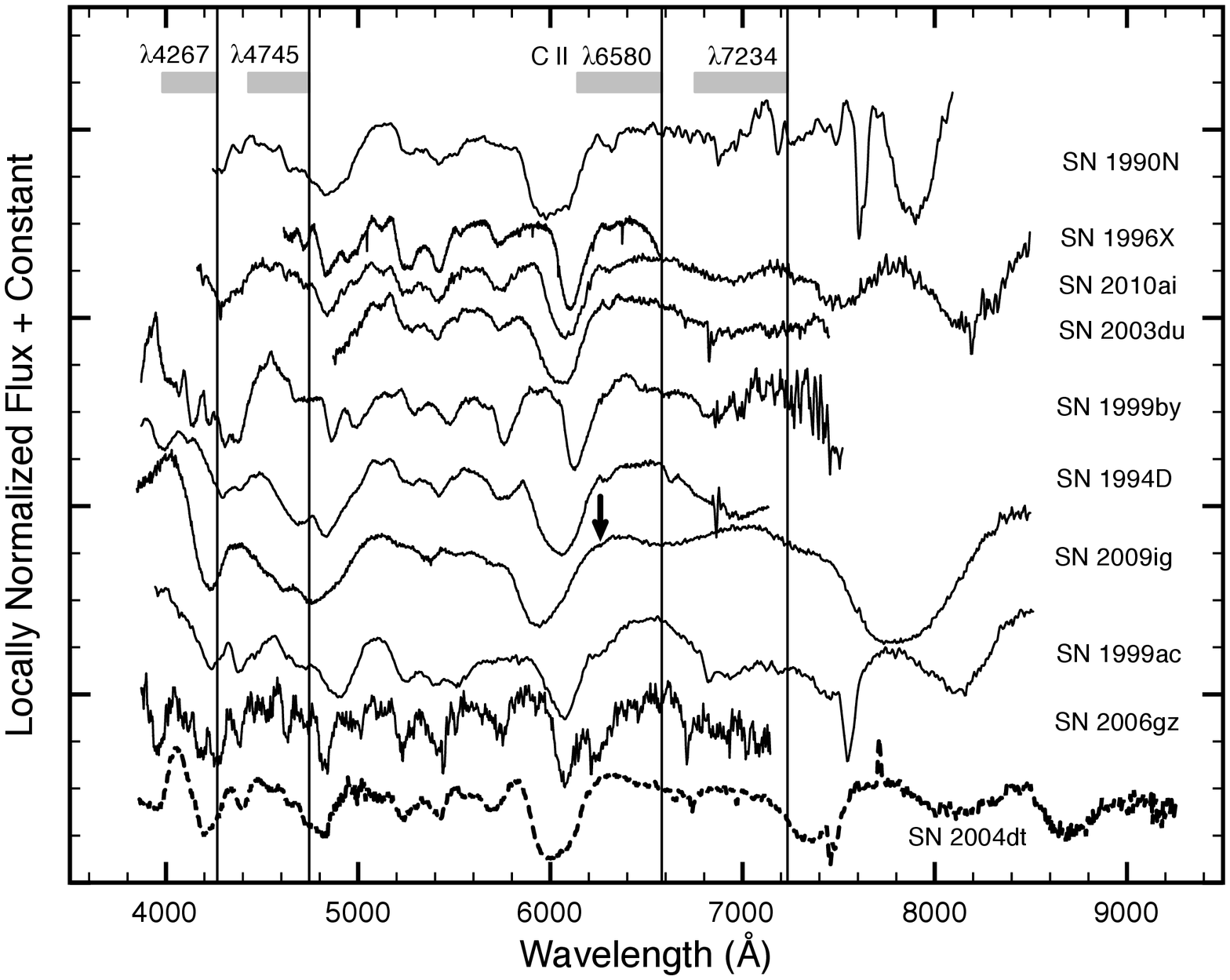}
\caption{Sample of 19 SNe Ia whose pre-maximum optical spectra contain an absorption due to \ion{C}{2} $\lambda$6580. A pre-maximum spectrum of SN 2004dt (HVG) is shown at the bottom (dashed line) for comparison. Spectra have been normalized and corrected for host galaxy redshift. Vertical lines indicate the rest wavelength position of \ion{C}{2} $\lambda\lambda$4267, 4745, 6580, and 7234. The width of the shaded regions near each vertical line corresponds to Doppler velocities between 1000 - 20,000 km s$^{-1}$. To guide the eye, a couple black arrows indicate the position of a likely absorption feature of \ion{C}{2} $\lambda$6580.}
\label{fig:sample}
\end{figure}

In Figure~\ref{fig:sample}, we present a single pre-maximum spectrum of the 19
SNe Ia in our sample that show evidence of a \ion{C}{2} $\lambda$6580
absorption signature. Under the assumption of local thermodynamic equilibrium
at 10,000 K, the strongest four optical lines of \ion{C}{2} are
$\lambda\lambda$4267, 4745, 6580, and 7234, with the 6580 \AA\ line being the
strongest. Rest frame positions of these lines are represented as vertical
lines in Figure~\ref{fig:sample} and the four grey bands at the top indicate
Doppler velocities of 1000$-$20,000 km s$^{-1}$. This corresponds to a region
spanning 440 \AA\ wide and blueward of 6580 \AA, well within the extent of typical
\ion{Si}{2} $\lambda$6355 P-Cygni profiles. 

In Table 2 we also note the epoch of the earliest spectrum taken and how likely
a \ion{C}{2} $\lambda$6580 detection is for each SN Ia (see \S5.2). If both
\ion{C}{2} $\lambda$6580 and $\lambda$7234 absorption features are present or
if the \ion{C}{2} $\lambda$6580 is an obvious notch, then we denote the
detection as ``Definite''. If a weaker notch is seen, and both absorption wings
remain, then the detection is labeled as ``Probable''. If a $\sim$6300 \AA\
feature comes in the form of a ``slump'' on the emission component of
\ion{Si}{2} $\lambda$6355, then we note the \ion{C}{2} $\lambda$6580 detection
as ``Possible''. Some of the published data have S/N ratios that prevent making
a clear case for \ion{C}{2} $\lambda$6580 detection and we note these as
``Uncertain''. If none of these criteria are met, then there is ``No''
detection. 

\section{SNe Ia: Subtype Classes}

The observed diversity of SNe Ia has been subdivided based on photometric and
spectroscopic properties by \cite{Benetti05} and \cite{Branch06}. Below we
briefly describe these classification schemes which will be referred to in our
analysis.

\cite{Benetti05} grouped 26 SNe Ia according to two photometric and three
spectroscopic observables, namely (1) $\Delta$$m$$_{15}(B)$ $-$ the decline in
magnitude of the B-band 15 days after B-band maximum \citep{Phillips93}, (2)
$M_{B}$ $-$ the peak B-band magnitude, (3) $\dot{v}_{Si II}$ $-$ the \ion{Si}{2}
expansion velocity rate of decrease, (4) $v_{10}$(\ion{Si}{2}) $-$ the
expansion velocity of \ion{Si}{2} $\lambda$6355 10 days after maximum light,
and (5) ${\cal R}(Si)_{max}$ $-$ the ratio of $\lambda\lambda$5972, 6355
absorption depth measured at maximum light \citep{Nugent95,Bongard08}. 

With a sample of 26 SNe Ia, they found that SNe Ia could be organized into
three discrete subtypes that were functions of velocity gradient and
luminosity, namely (1) high-velocity gradient (HVG), (2) low-velocity gradient
(LVG), and (3) FAINT.  The subtypes HVG and LVG were mainly distinguished by
having $\dot{v}$ values above or below $\sim$70 km s$^{-1}$ day$^{-1}$,
respectively, while the SNe in their sample with $M_{B, max}$ $\gtrapprox$
$-$18.20 were labeled as FAINT. 

\cite{Branch06} took a purely spectroscopically based approach, classifying 24
SNe Ia (later 65 SNe; see \citealt{Branch09}) by the equivalent width of
features near 5750 \AA\ and 6100 \AA\ which are usually attributed to
\ion{Si}{2} $\lambda\lambda$5972, 6355, respectively. When SNe Ia are arranged
in this manner, the objects can be subdivided roughly into four spectroscopic
subtypes: (1) Core-Normal (CN) SNe Ia consist of objects such as SN 1994D,
where from pre-maximum to 1-week post-maximum spectra are dominated by lines of
\ion{Ca}{2}, \ion{Fe}{2}, \ion{Fe}{3}, \ion{Mg}{2}, \ion{O}{1}, \ion{S}{2},
\ion{Si}{2}, and \ion{Si}{3},  (2) Broad-Line (BL) SNe Ia are similar to CNs
but instead display noticeably broader lines with higher average Doppler
velocities, (3) Cool (CL) 1991bg-like spectra which exhibit low ionization energy
ions such as \ion{Ti}{2}, along with an increased ratio between the 5750 and
6100 \AA\ features, and (4) Shallow Silicon (SS) 1991T-likes display mostly
high ionization energy ions, such as \ion{Fe}{3}, in their pre-maximum spectra
and are accompanied by weak \ion{S}{2} absorption features. 

We note that comparing the sample of SNe Ia used in both studies shows that
FAINT and HVG objects are equivalent to CL and BL respectively, and the LVG
objects are equivalent to CN and SS subtypes \citep{Branch06}. Since we are
using subtype classes that are mostly based on spectroscopic properties of SNe
Ia, our study cannot address issues regarding broadband photometric
observations. A systematic study of both the spectroscopic and light-curve
properties is difficult because these data can come from multiple sources and not
every target we study here have both good spectroscopy and photometry. This
type of study is better suited for the Nearby Supernova Factory
\citep{Aldering02} and the Palomar Transient Factory \citep{Rau09}, where the
data sets are large and have good time-series coverage.

\section{Spectrum Analysis Model: \texttt{SYNOW}}

One way to infer the velocity range of unburned material is to measure the
absorption minimum of the \ion{C}{2} $\lambda$6580 line. The observed minimum
of this feature is often located near the strong emission component of
\ion{Si}{2} $\lambda$6355, and is therefore subject to line blending and any
limb-brightening that the \ion{Si}{2} line contributes to the integrated
spectrum \citep{Hoflich90}. Consequently, the observed minimum may
underestimate the actual expansion velocity of the carbon-rich region.
Therefore, in order to accurately estimate the true minimum of a blended line
profile, one must take into account the effects of line formation by
reconstructing the spectrum via numerical calculation.

While there have been several new and more detailed spectrum synthesis codes
presented in the literature since the inception of \texttt{SYNOW} nearly thirty
years ago, the \texttt{SYNOW} spectrum synthesis model
remains a useful tool for the quick analysis of resonant scattering line
profiles \citep{Baron94,Mazzali00,Branch07,Kasen08,Baron09}. Therefore, we
chose the less computationally intensive approach of \texttt{SYNOW} to produce
results that were internally consistent when we compared the SNe Ia in our
sample. 

From a single-epoch optical spectrum, one can reproduce many of the conspicuous
features seen in SNe of all subtypes using \texttt{SYNOW}. However, to have
greater confidence in the identification of a spectroscopic feature, it is best
to have a time series of closely spaced spectra in order to follow the
evolution of both the observations and the fit. Because we wish to probe the
nature of \ion{C}{2} $\lambda$6580 features in SNe Ia, we required fitting all
spectra for the duration of time in which the 6300 \AA\ features could be seen. 

In \texttt{SYNOW}, one computes a spectrum by specifying the location and
optical depth for a given set of ions. This allows one to infer spectral line
identifications by directly fitting to a series of observed spectra. While
\texttt{SYNOW} does not calculate relative abundances of various elements, it
is instructive at reconstructing the complex spectroscopic profiles brought
about by multiple line scattering that is inherent to moving media. 

The version of the \texttt{SYNOW} model that we used can be described as
follows: (1) a spherically symmetric and homologously expanding ejecta is
modeled using a v $\propto$ r law, (2) light is emitted from a sharp
photosphere,  (3) optical depth, $\tau$, is a function of velocity as either
$(v/v_{phot})^{-n}$, $exp^{-\frac{v-v_{phot}}{v_{e}}}$, or
$exp^{-\frac{(v-v_{phot})^{2}}{2\sigma^{2}}}$ where each of these functions are
characterized by indices n, v$_{e}$, and $\sigma$ respectively, (4) line
formation is purely due to resonant scattering and is treated using the Sobolev
Approximation \citep{Sobolev57,Jeffery90}, and (5) for a given ion a reference
line profile is calculated for a given $\tau$ and the remaining lines follow
from Boltzmann statistics. Input parameters for a \texttt{SYNOW} spectral fit
include: (i) a photospheric velocity (v$_{phot}$), (ii) reference line $\tau$
and minimum/maximum velocities for each ion, and (iii) excitation temperature,
$T_{exc}$, to determine LTE level populations with respect to a reference line.  

As was done in the SN Ia comparative study of \cite{Branch05}, 
for our invesitigation we too left the excitation temperature at 10,000 K for each ion. 
We also chose to use the exponential form of 
the optical depth profile with v$_{e}$ = 1000 km s$^{-1}$. 
This aided in limiting the number of free parameters for a spectral fit. 

Because \texttt{SYNOW} assumes spherical symmetry, we are limited when
investigating the possible asymmetrical distributions of unburned material. If
the distribution is that of a spherical layer, as is the case for the W7 model
\citep{Nomoto84, TNY86}, then it is straightforward to make \texttt{SYNOW}
model fits to compare with observations. In this case, the unburned material
resides above the burned material in a spherical shell where the thickness
depends on the extent of the burning. For W7, this boundary is roughly at
14,000 km s$^{-1}$ with a stratified composition of IMEs and IPEs below. 

Recent multidimensional models of delayed detonations suggest that
the unburned material may be left behind in clumps throughout the ejecta
\citep{Gamezo04}. For the situation where the unburned material is heavily
concentrated to a single clump structure, we can only utilize \texttt{SYNOW} in
certain cases (see \S5.4). 

Observations have shown that there often exists higher velocity regions of line
formation; namely \ion{Ca}{2} and \ion{Si}{2}
\citep{Hatano99a,Kasen03,Tanaka09}. Such regions are said to be $\mathit{detached}$
from the photosphere. We note a noticeable facet that arises when fitting
\ion{Si}{2} $\lambda$6355 profiles during the earliest epochs is that the
absorption width, the slow rise of the blue wing, and the sharp rise of the red
wing require two separate velocity components of \ion{Si}{2} to achieve a good
match to observations. At best, a single-component of \ion{Si}{2} with an
increased value of the optical depth profile indices can only properly fit the
red wing of the 6355 \AA\ absorption feature. 

G. H. Marion et al. (2011, in preparation) discuss recent observations of the Type Ia event, SN
2009ig, where there is clear evidence for both a photospheric region and a high
velocity region of \ion{Si}{2}. In their \texttt{SYNOW} analysis, they used the
two-component approach in following the evolution of the \ion{Si}{2}
$\lambda$6355 feature, which produced a better fit overall. Similarly, we too
have adopted a procedure of using two components of \ion{Si}{2} that are
separated by $\sim$4000$-$6000 km s$^{-1}$ when necessary. 

One consequence that detached ions have on the line profile is that the
emission component of the line is flat-topped. Some of our spectroscopic fits
in \S5.3 detach \ion{Si}{2} to better fit the absorption component, while
forfeiting a comparable fit to the full emission component; e.g. our fit for SN
1999ac. The impact that detaching has towards blending with \ion{C}{2}
$\lambda$6580 is minimal and only one of offsetting the prescribed value of
$\tau$ for \ion{C}{2} in the fit. 

\section{Results}

\begin{figure}
\includegraphics[width=\linewidth]{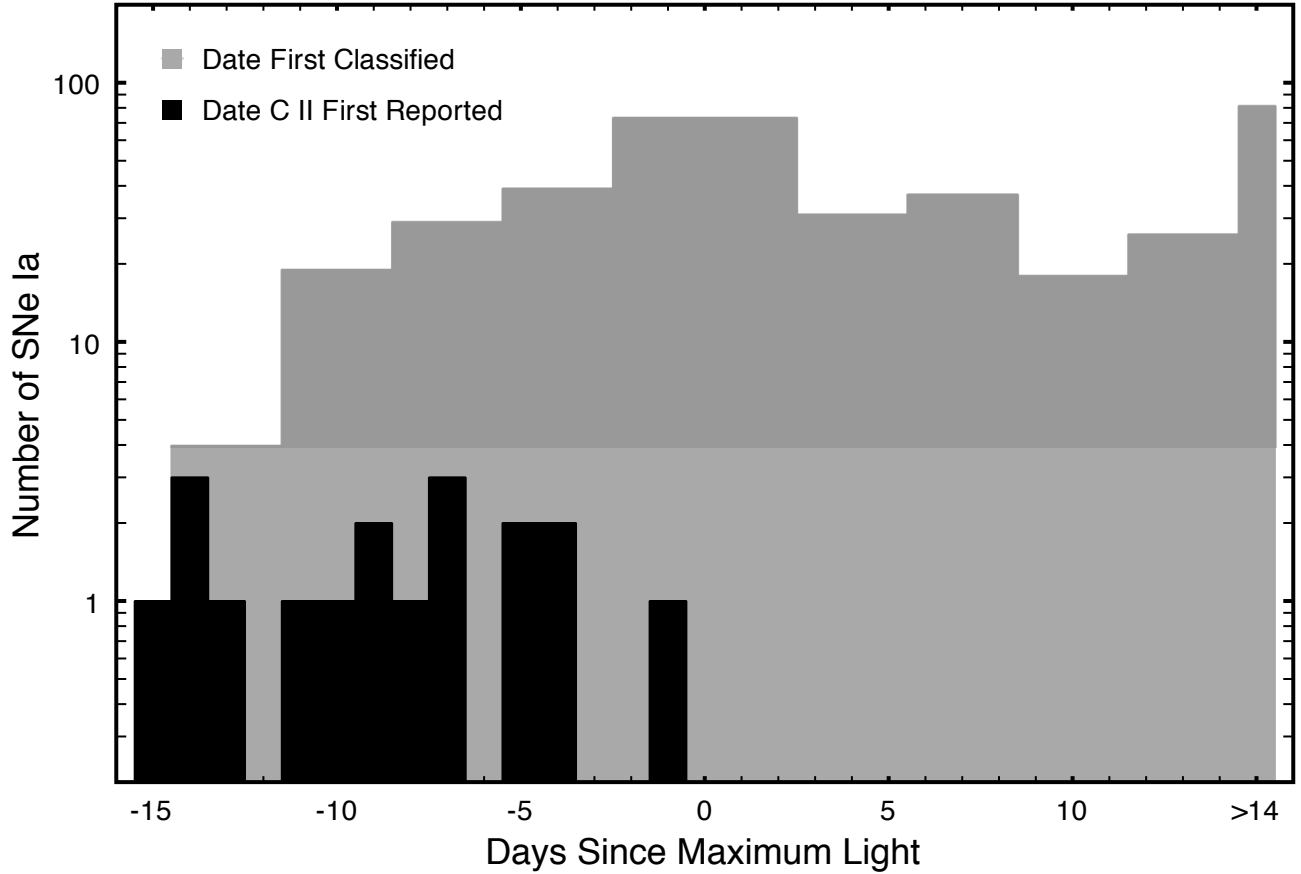}
\caption{Shaded in grey are the number of SNe Ia
discovered between 2006 January 1 and 2009 December 1 plotted against the epoch at which they were identified as being a Type Ia supernovae. Shaded in black are the number of SNe Ia that show $\sim$6300 absorption features in their pre-maximum spectra at the time when the feature was first seen. Data were taken from Central Bureau Electronic Telegrams (CBET) and Astronomical Telegrams (ATEL). Estimates for the age of each supernova were obtained by one of several publicly available spectrum-comparison tools (PASSparToo; \cite{Harutyunyan05}, Superfit; \cite{Howell05}, SNID; \cite{BT07}, GELATO; \cite{Harutyunyan08}). The cited epoch for each SNe Ia in this plot was taken as reported in the telegrams. The numbers above are not a strict sampling of all objects during the four year period, given that some of the age estimates were too vague to be included. This explains the peaks at day $-7$, 0, $+$7 and $+$14 relative to maximum light. Because the spectral identification programs determine age with several days of error, we binned the data sample at days $-13\pm1$, $-10\pm1$, $-7\pm1$, $-4\pm1$, $0\pm2$, $+4\pm1$, $+7\pm1$, $+10\pm1$ and $+13\pm1$. Both histograms show, on average, that observations before 1$-$week pre-maximum are not well represented compared to observations near maximum light or later.}
\label{fig:histogram}
\end{figure}

In terms of the spectroscopic diversity of SNe Ia, many of the differences are seen
immediately after the explosion as the photosphere maps out the distribution of
the outermost ejected material, with an increase toward spectroscopic
conformity at later epochs \citep{Branch08}. Unfortunately, as shown in
Figure~\ref{fig:histogram}, SN Ia spectroscopic observations
during the two weeks prior to maximum light are underrepresented compared to
those taken near maximum light or at later times. In addition,
absorption features attributed to \ion{C}{2} $\lambda$6580 are generally weak.
As a result, they can be easily missed and thus go unreported or are
not securely verifiable due to low S/N and/or line blending.

\subsection{Frequency of \ion{C}{2} Absorption Features}

Our search of the literature revealed that $\sim$30\% of SNe Ia with
moderate to high S/N, pre-maximum spectra taken since 1983 January 1 show a feature near
6300 \AA\ that may be associated with \ion{C}{2} $\lambda$6580 absorption (see Table 2; Definite + Probable). Because of the sparsity of optical spectra before peak brightness, this percentage may not represent the actual fraction of SNe Ia that exhibit \ion{C}{2} $\lambda$6580 signatures during the pre-maximum phase.  Furthermore, the amount of published SN Ia data does not equal the amount of data taken, thereby placing considerable uncertainty on such an estimate. 

For comparison, in Figure~\ref{fig:histogram}, we also tally and plot the total number of SNe Ia where \ion{C}{2} $\lambda$6580 was first detected with respect to maximum light. When the total number of published \ion{C}{2} observations over the past couple decades is compared to the total number of SNe Ia discovered during the pre-maximum phase in a span of three years, the occurrence of absorption due to \ion{C}{2} $\lambda$6580 appears infrequent. 

However, many SNe Ia might show \ion{C}{2} (or \ion{C}{1}, or \ion{C}{3}; see
\S6.1) during the first weeks of the explosion, but may not exhibit strong,
identifiable carbon absorption features by the time they are discovered and
spectra taken.  If spectra are not obtained prior to maximum light, this can
lead to an under-reporting of \ion{C}{2} SNe Ia cases.  For example, in some
objects, the \ion{C}{2} features appear to fade by day $-$5 or even earlier.
This raises the question: Might all SNe Ia show appreciable \ion{C}{2}
absorption at some level if observed at a sufficiently early enough epoch? 

\begin{figure}
\includegraphics[width=\linewidth]{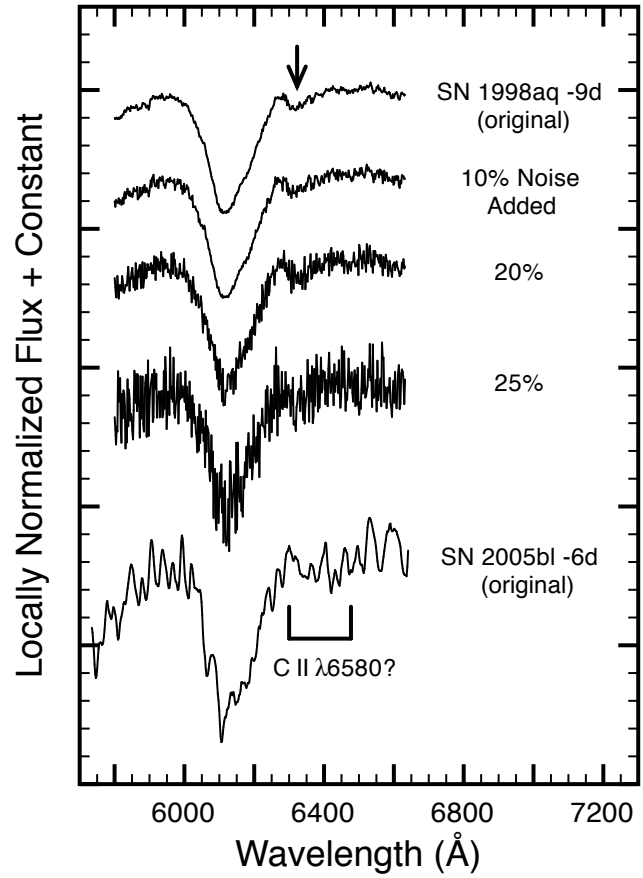}
\caption{Five segments of spectra showing the questionable of a \ion{C}{2} $\lambda$6580 absorption feature as a function of the signal-to-noise ratio. At the top is the original spectrum for the day $-$9 observation of SN 1998aq. The three spectra that follow represent the day $-$9 spectrum with 10, 20, and 25\% Gaussian noise added. The spectrum at the bottom is that of SN 2005bl, six days before maximum light, where \cite{Taubenberger08} identified the 6400 \AA\ feature with that of \ion{C}{2} $\lambda$6580.}
\label{fig:noise}
\end{figure}

Given the weak impact that \ion{C}{2} features are observed to have on the
integrated optical spectrum, identifying every instance for when \ion{C}{2} is
present rests heavily on one's ability to distinguish the signature from noise
and the effects of line blending. In Figure~\ref{fig:noise}, we compare the
spectrum of SN 1998aq at day $-$9 to three cases where we have artificially
added Gaussian noise to the original spectrum. With only 10\% and 20\% noise
added, the \ion{C}{2} $\lambda$6580 feature remains distinguishable. However,
at 25\% the feature begins to lose its identity above the noise. An example of
this is the spectrum of SN 2005bl at day $-$6, shown in Figure~\ref{fig:noise}.
\cite{Taubenberger08} proposed a \ion{C}{2} $\lambda$6580 identification for
the 6400 \AA\ depression seen in the spectrum. However, we excluded this SN
from our sample since we were not able to generate an accurate fit for the
feature.

\subsection{SNe Ia \ion{C}{2} Features: Conspicuous to Weak}

Like most features in the spectra of SNe Ia, the degree of adjacent line
blending that \ion{C}{2} $\lambda$6580 absorption signatures undergo can
obscure the full extent of a line profile. More often than not, the minimum of
the \ion{C}{2} $\lambda$6580 line is blended with the P-Cygni emission
component of the \ion{Si}{2} $\lambda$6355 line. The day $-$5 spectrum of SN
1999by is a good example of when this takes place (see
Figure~\ref{fig:sample}). 

When the \ion{C}{2} $\lambda$6580 absorption feature is weak there is also often no
obvious \ion{C}{2} $\lambda$7234 signature. However, when the \ion{C}{2}
$\lambda$6580 feature is strong, a corresponding \ion{C}{2} $\lambda$7234
feature does begin to appear as the excited level of the 7234 \AA\ line becomes
more populated. There are several cases, such as SN 2006gz and 2007if, where
\ion{C}{2} $\lambda$6580 is certainly present but either the wavelength
coverage is incomplete or the spectrum is too noisy to say whether or not
\ion{C}{2} $\lambda$7234 is present as well. 

\begin{figure}
\includegraphics[width=\linewidth]{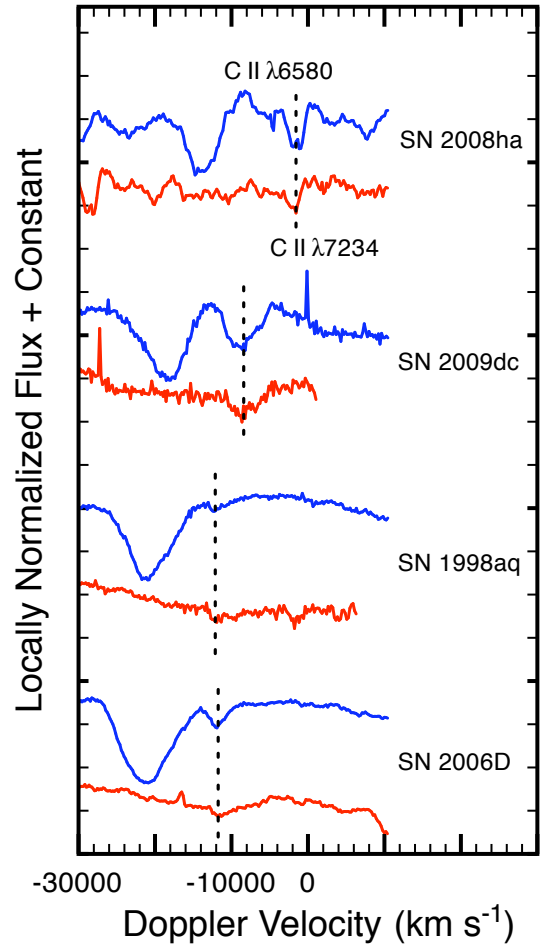}
\caption{Doppler velocity scaled spectra for four of the SNe Ia in our sample. Blue and red lines represent \ion{C}{2} $\lambda$6580, 7234, respectively, for each SN. Dashed vertical lines are positioned at the minima of the absorption features to guide the eye.}
\label{fig:overlap}
\end{figure}

In Figure~\ref{fig:overlap}, we plot two spectral regions of four SNe Ia in
Figure~\ref{fig:sample} that highlight cases when both of these \ion{C}{2} lines are
present. For each SN, the red and blue lines denote $\lambda\lambda$6580, 7234
Doppler velocity-scaled spectra, respectively. The absorption minima overlap
nicely (see dashed vertical lines) and the symmetry about the minima is
indicative of a match between the spectral signatures of the same ion. We
interpret this to mean that the 7234 \AA\ line is indeed present when the 6580
\AA\ line is strong.  Currently, these two \ion{C}{2} lines are the best means
by which to securely identify whether carbon is present in the early-epoch SN Ia
spectra. 

The \ion{C}{2} lines that appear further in the blue have also been suggested
to be present as well \citep{Thomas07}. However, this region is crowded by
lines of IPEs making it more difficult to securely identify \ion{C}{2}
$\lambda\lambda$4267, 4745 and use for determining accurate Doppler velocities.
This appears to even be the case for SN 2006gz, 2007if, and 2009dc where the
\ion{C}{2} $\lambda$6580 absorption is strong \citep{Hicken07,Scalzo10,Taubenberger10}. 

\subsection{\texttt{SYNOW} Model Fitting}

For each of the 19 SNe Ia in our \ion{C}{2} sample, we produced a time series of synthetic optical spectra that covered the observed extent of the \ion{C}{2} $\lambda$6580 feature's presence as well as the observed wavelength coverage of the data ($\sim$3500$-$9000 \AA). Our initial fits included the canonical set of IMEs and IPEs that are prevalent in pre-maximum SN Ia spectra, whereby afterward we included \ion{C}{2} and adjusted the optical depth and detachment velocity until a match to observation was made. The goodness-of-fit to the observed \ion{Si}{2} $\lambda$6355 $-$ \ion{C}{2} $\lambda$6580 blended profiles did not change with the full set of IMEs and IPEs removed. Thus, for each SN in Table 3, we only list the relevant parameters for \ion{C}{2} and \ion{Si}{2}. 

As was discussed above in \S4, some of the observed \ion{Si}{2} $\lambda$6355 profiles required two separate components of \ion{Si}{2} in order to fit the full width of the absorption. We indicate which SNe Ia in our sample required this fitting procedure by appending the inferred Doppler velocities and optical depths with ``$>$'' in Table 3.

\begin{figure}
\includegraphics[width=14cm]{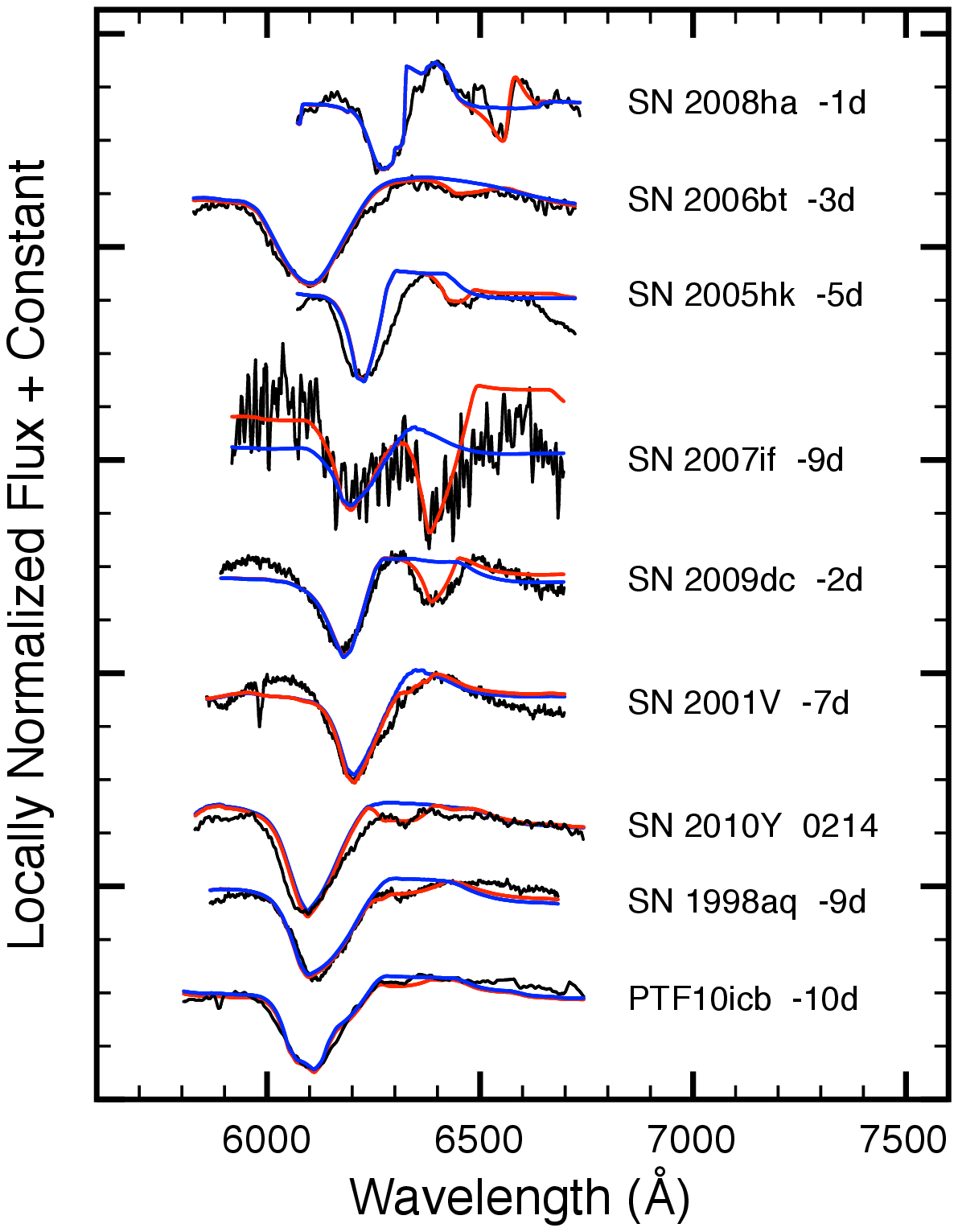}
\includegraphics[width=14cm]{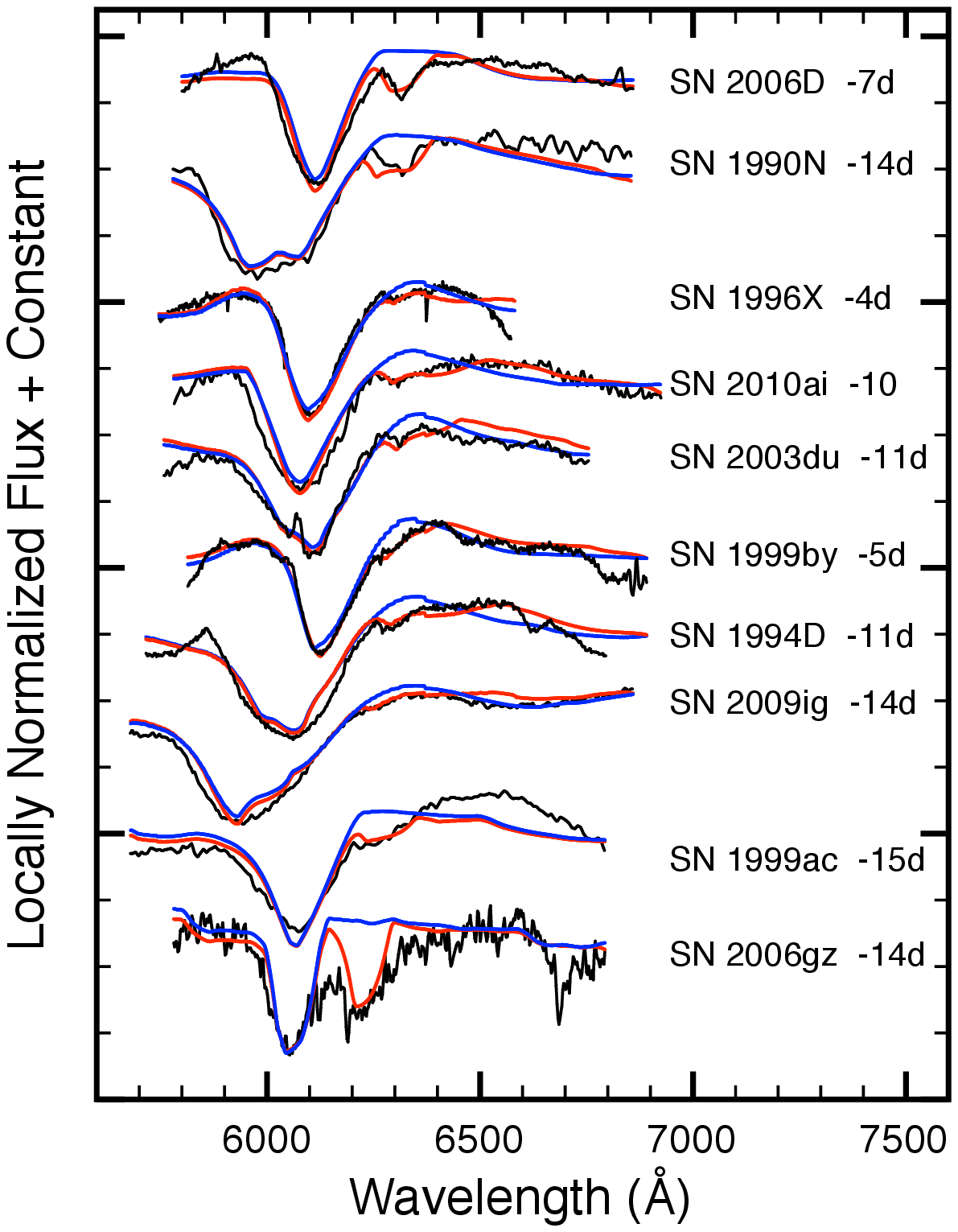}
\caption{\texttt{SYNOW} fits of the 19 SNe Ia in our sample presented in two panels. The red and blue lines denote a synthetic fit to the spectrum of a SNe Ia with and without C II, respectively. The contrast between both fits suggests the presence of \ion{C}{2} with varying strength in these objects. See Table 2 for fitting parameters.}
\label{fig:samplefits}
\end{figure}

In Figure~\ref{fig:samplefits}, we compare fits to a single observed spectrum for each of the SNe Ia in our sample. The black lines represent the observed spectra, the red lines are \texttt{SYNOW} fits where \ion{C}{2} has been included, and the blue lines are the same fits without \ion{C}{2}. The synthetic spectra match fairly well to a variety of \ion{Si}{2} $\lambda$6355 $-$ \ion{C}{2} $\lambda$6580 blended profiles, and the interpretation that the 6300 \AA\ feature is due to \ion{C}{2} $\lambda$6580 is in agreement with that of previous authors (see references in Table 2). 

In this study, we also offer new and revised expansion velocity estimates of \ion{C}{2} for a couple SNe Ia, particularly SN 1990N and 1999by. It was suggested by \cite{Fisher97} that the overly broadened 6040 \AA\ absorption in the day $-$14 spectrum of SN 1990N was due to a two-component blend composed of \ion{Si}{2} at $\sim$20,000 km s$^{-1}$ and \ion{C}{2} at $\sim$26,000 km s$^{-1}$. Similarly, \cite{Mazzali01} suggested that this feature is predominantly due to \ion{Si}{2} $\lambda$6355 but also requires an outer zone of high velocity carbon between 19,000 and 30,000 km s$^{-1}$. If this interpretation is correct, then other SNe Ia like SN 1990N would also require a similar zone of carbon to reproduce their early-epoch spectra. However, our fit for SN 1990N  uses two components of \ion{Si}{2} to fill the 6040 \AA\ feature while \ion{C}{2} is only at 16,000 km s$^{-1}$ to account for the 6300 \AA\ feature.

In the case of the sub-luminous SN 1999by, \ion{C}{2} $\lambda$6580 was not reported by \cite{Garnavich04}. However, our fit in Figure~\ref{fig:samplefits} is fairly convincing when \ion{C}{2} is included as a detached layer 2000 km s$^{-1}$ above the photosphere for the day $-$5 spectrum. In fact, almost 80\% of the SNe in our sample suggest at least a mildly detached layer of \ion{C}{2} during some point along the evolution of the absorption feature. For example, we modeled the \ion{C}{2} in SN 1994D initially at 14,000 km s$^{-1}$ coincident with $v_{phot}$ at day $-$11, after which it remains at this velocity as the photosphere recedes to 12,000 km s$^{-1}$ by day $-$5. 

\section{Discussion}

The strength and velocity range of carbon in pre-maximum SNe Ia spectra
can provide a valuable tool to investigate various explosion models. The W7 deflagration model, for instance, contains a $\sim$0.07$M_{\odot}$ layer of unburned material above 14,000 km s$^{-1}$. Using the spectrum synthesis and model atmosphere code, \texttt{PHOENIX}, \cite{Lentz01} compared calculated non-LTE spectra to spectroscopic observations of the normal Ia event, SN 1994D. Despite the outer layer of unburned material in W7, none of their synthetic spectra were able to account for the observed \ion{C}{2} $\lambda$6580 absorption feature near 6290 \AA. Our \texttt{SYNOW} fits for this object allow for \ion{C}{2} below 14,000 km s$^{-1}$. To that end, \cite{Tanaka10} were able to reproduce a \ion{C}{2} $\lambda$6580 absorption feature seen in the day $-$11 spectrum of SN 2003du by placing the carbon-rich region at lower velocities in their version of W7. They obtained an upper limit on the
abundance of carbon to be 0.016$M_{\odot}$ at $v$ $>$ 10,500 km s$^{-1}$ in this object. 

At late-times and based on the three-dimensional deflagration models of
\cite{Roepke05}, \cite{Kozma05} compared synthetic spectra to the late-time
spectra of three SNe Ia and set an upper mass limit of unburned material below
10,000 km s$^{-1}$ to be $\sim$0.07 $M_{\odot}$. Other estimates for the mass
of unburned material have been made using delayed-detonations
\citep{Hoeflich02} and other modeling; e.g. lower limit of 0.014$M_{\odot}$ of
unburned material between 10,000 and 14,000 km s$^{-1}$ for SN 2006D
\citep{Thomas07}. However, since each estimate is obtained by different means,
a comparison of such results does not advance  
the discussion on the nature of \ion{C}{2} $\lambda$6580 absorption features. 

In order to utilize \ion{C}{2} $\lambda$6580 absorption features for estimating the mass of unburned material, the effects of temperature and geometry of the carbon-rich regions as well as the influence of radiative transfer effects must be explored. Our \texttt{SYNOW} modeling begins this process by mapping the observed velocity distribution for a sample of 19 SNe Ia. Below, we discuss our interpretation of the observed frequency of \ion{C}{2} absorption features and how they relate to the properties of carbon-rich regions and SN Ia diversity. 

\subsection{Temperature Effects on Carbon Features}

The temperature of the carbon-rich region will influence whether or not
\ion{C}{2} is the dominant ionization species and will therefore dictate the
strength of \ion{C}{2} $\lambda$6580 absorption features. Using non-LTE
calculated spectra, \cite{Nugent95} pointed out that the spectroscopic sequence
observed among various SNe Ia could be explained by a continuous change in the
effective temperature of the ejecta (7400$-$11,000 K), from cool 1991bg-likes to
the hotter 1991T-likes (see their Fig.\ 1). While the ejecta of SNe Ia are an
environment with non-LTE processes, under the assumption of a C+O-rich
composition the Saha-Boltzmann equation indicates that carbon will mostly
be in the form of \ion{C}{2} between 6000 and 12,000 K.

Of the 68 SNe Ia listed in Table 2, 10 out of the 14 CNs and five out of the 13
CLs exhibit \ion{C}{2} $\lambda$6580 absorption features. From a nearly equal
sampling of SN Ia subtypes, the fact that two CNs for every one CL SN Ia
exhibit \ion{C}{2} absorption lines suggests that the presence of \ion{C}{2}
absorption features depend on the effective temperature to some degree. This
point is also consistent with the fact that only two of 13 SSs show signs of
\ion{C}{2} $\lambda$6580 absorption features in their spectra. 

We note that \ion{C}{1} absorption features are not often detected in SN Ia
spectra. \cite{Marion06} presented NIR spectra of three normal SNe Ia and
discussed the lack of \ion{C}{1} absorption signatures due to the absence of
the strongest \ion{C}{1} NIR lines, namely \ion{C}{1} $\lambda\lambda$9093,
10691. A more extensive study of 41 SN Ia NIR spectra spanning two weeks before
and after maximum light  was discussed by \cite{Marion09} and they too reported
a lack of \ion{C}{1} signatures. Because spectral signatures of carbon burning
products were observed to occupy the same region of ejecta (\ion{Mg}{2} and
\ion{O}{1}), \cite{Marion06} concluded that nuclear burning had been complete
out to at least 18,000 km s$^{-1}$ in their objects. 

It is perhaps not surprising that many NIR spectra of SNe Ia do not show lines
from \ion{C}{1}, given that (1) the abundance of carbon may be only
$\sim$1-10\% of the total ejected mass, and (2) if there is carbon present in
the outer layers then it is mostly once ionized. Interestingly, however,
\cite{Hoeflich02} reported a conspicuous \ion{C}{1} $\lambda$10691 absorption
feature in the day $-$4 NIR spectrum of SN 1999by. They were able to reproduce
many features of the observed optical and NIR spectra using a series of
sub-luminous delayed-detonation models with a range of transition densities,
$\rho_{tr}$, between 8$-$27 $\times$ 10$^{6}$ g cm$^{-3}$. 

The appearance of the \ion{C}{1} $\lambda$10691 line was seen concurrently with
the optical \ion{C}{2} $\lambda$6580 absorption feature we reported above in
\S5.3. Both the \ion{C}{1} and \ion{C}{2} spectral signatures indicated that
the carbon-rich region was above the 12,000 km s$^{-1}$ photosphere; i.e. the
minimum of the \ion{C}{1} $\lambda$10691 absorption corresponds to $\sim$13,000
km s$^{-1}$ while our \texttt{SYNOW} fits for this object place \ion{C}{2} at
14,500 km s$^{-1}$. At least for this cool sub-luminous SNe Ia, the influence
of temperature on the ionization state of the carbon-rich region is apparent
from the simultaneous appearance of \ion{C}{1} and \ion{C}{2} absorption
features.  

Additional evidence that ejecta temperature plays a role in the detection of
carbon in SN Ia spectra would be if \ion{C}{3} were clearly detected in a
hotter SN Ia subtype, such as SN 1991T or SN 1997br. The similarity between the
ionization potentials of \ion{C}{2} (24.4 eV) and \ion{S}{2} (23.3 eV) suggests
that the presence of \ion{C}{3} absorption features may be concurrent with
spectroscopic signatures of \ion{S}{3}. Fortunately, we can examine optical
spectra to check for the simultaneous presence of \ion{C}{3} $\lambda$4649 and
\ion{S}{3} $\lambda$4254 absorption features. 

\begin{figure}
\includegraphics[width=\linewidth]{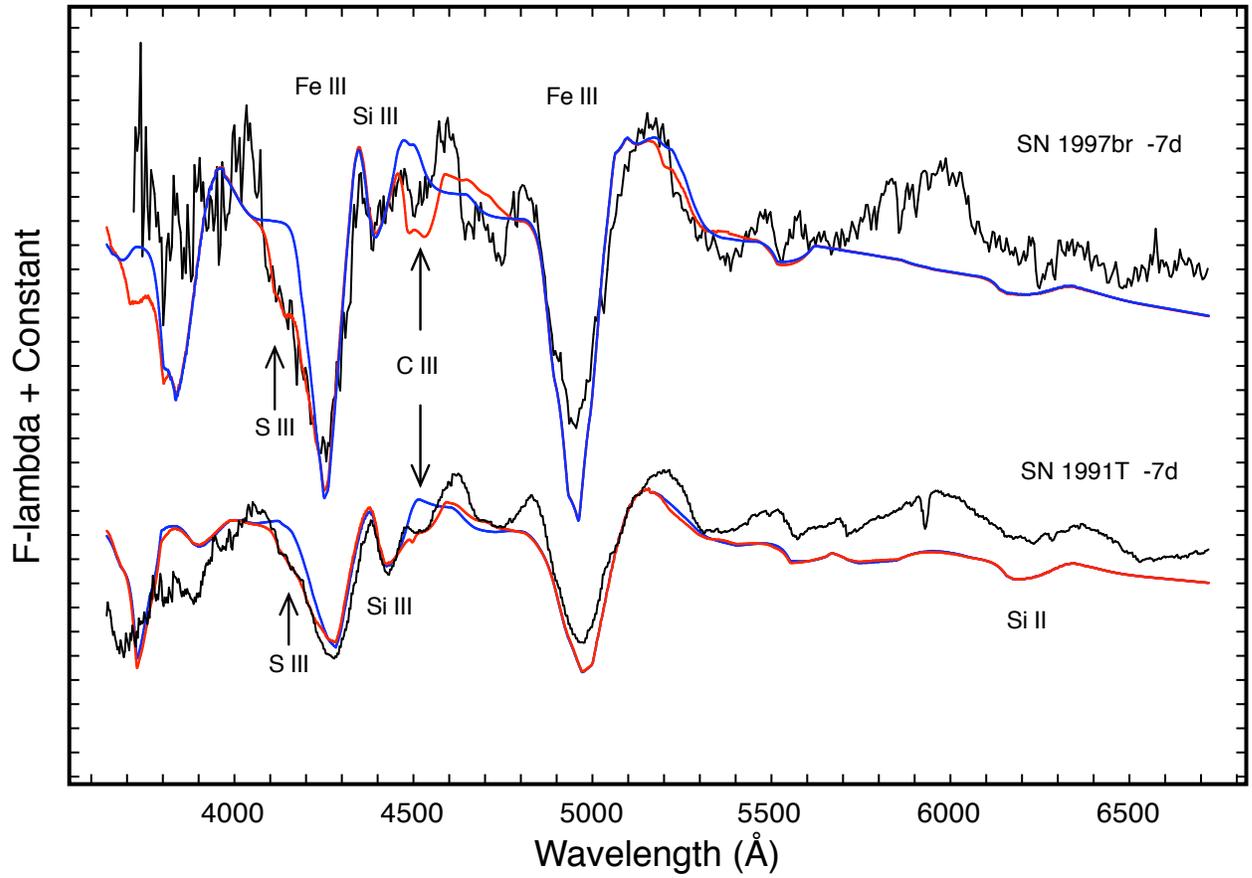}
\caption{\texttt{SYNOW} fits for SN 1991T and 1997br, two SS SNe Ia, around 1-week pre-maximum. In addition to matching the major features with \ion{Fe}{3}, \ion{S}{2}, \ion{Si}{2}, and \ion{Si}{3}, the absorption at 4500 \AA\ is identified with a multiplet of \ion{C}{3} $\lambda$4649. As before in Figure~\ref{fig:samplefits}, the red line denotes the synthetic spectrum with carbon, and the blue line is without.}
\label{fig:ciii}
\end{figure}

In Figure~\ref{fig:ciii}, we plot and compare \texttt{SYNOW} fits for SN 1991T
and 1997br where we have included \ion{C}{3} and \ion{S}{3}. The identification
of the 4500 \AA\ feature of SN 1991T has been discussed before by
\cite{Hatano02}, and similarly in other SN Ia \citep{Garavini04,Chornock06}. While it was argued by \cite{Hatano02} that including \ion{C}{3} in the fit
produced a mis-match with the observed spectrum (too blue overall), our
synthetic spectra (red lines) are in fair agreement with observations near 4500
\AA\ for both SN 1991T and 1997br. 

In addition, a better fit to the 4250 \AA\ absorption feature is obtained with
the inclusion of \ion{S}{3}. The 4250 \AA\ feature is predominately due to the
\ion{Fe}{3} $\lambda$4404 multiplet. However, by adding \ion{S}{3} to the fit we were
able to fill in the blue wing of this absorption feature for both objects.
Therefore, our identification for the 4500 \AA\ absorption feature as being
that of \ion{C}{3} $\lambda$4649 is more likely, though the evidence is circumstantial. If the \ion{C}{3}
$\lambda$4649 identification is correct, then this may indicate that a lack of
\ion{C}{2} (or \ion{C}{1}) spectroscopic features do not necessarily imply the
complete burning of carbon in the hotter subtype events. 

\subsection{Interpreting \ion{C}{2} $\lambda$6580 Doppler Velocities}

In this paper, our attention has been primarily focused on carbon-rich regions
in the outermost ejecta. The wide range of observed \ion{C}{2} Doppler
velocities among different SNe Ia suggests a large variation in the extent of
carbon burning, with some objects exhibiting $\mathit{high}$ velocity carbon while
in others the carbon is present at $\mathit{low}$\ velocities (see Table 3).
How high or low is usually in reference to the position of the carbon
cut-off seen in the W7 model ($\sim$14,000 km s$^{-1}$) instead of the kinetic
energy of the supernova itself.  

\subsubsection{Extent of Burning via \ion{C}{2} $\lambda$6580}

Given that the characteristic ejecta velocity is proportional to
$(E_{kin}/M_{ej})^{1/2}$ \citep{Arnett82}, a standardized way of looking at the
extent of burning is a more appropriate measure for interpreting the range of
observed carbon velocities. The 6355 \AA\ line of \ion{Si}{2} has been used as
an indicator of the photospheric velocity at early epochs
\citep{Jeffery90,Patat96}. Choosing \ion{Si}{2} $\lambda$6355 over other lines
alleviates any difficulty in obtaining Doppler velocities amid too much line
blending and allows for consistent time-coverage. This makes the absorption
minimum of \ion{Si}{2} $\lambda$6355 a good point of reference for
investigating the extent of burning via \ion{C}{2} $\lambda$6580 absorption
features before maximum light. 

\begin{figure}
\includegraphics[width=\linewidth]{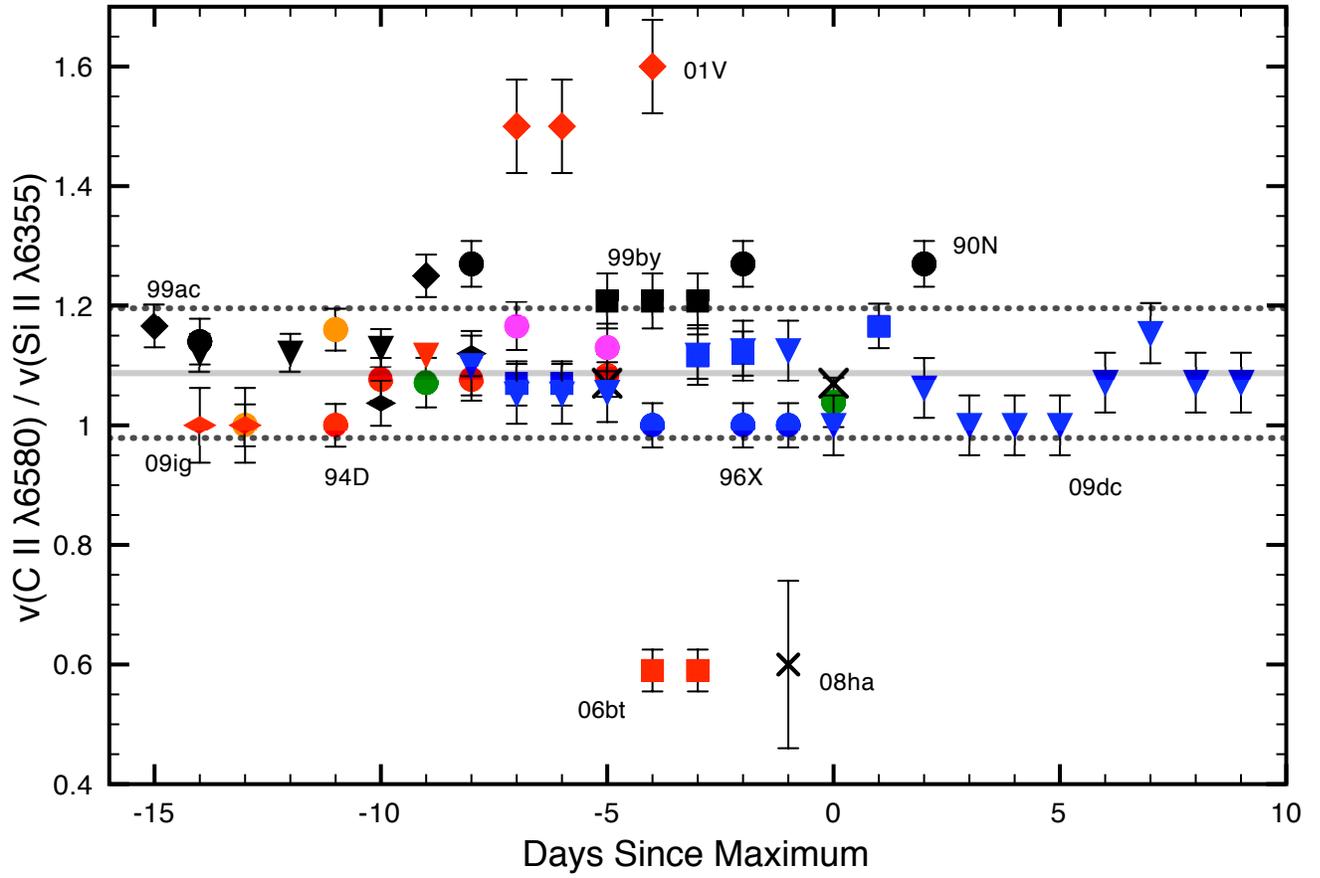}
\caption{$v$(\ion{C}{2} $\lambda$6580)/$v$(\ion{Si}{2} $\lambda$6355) vs. epoch for our \ion{C}{2}$-$SNe sample. Circles show CNs, side-ways diamonds show BLs, squares show CLs, diamonds show SSs, upside down triangles show Super-Chandra candidates, and the 'X' denotes the 2002cx-like, SN 2008ha. Calculated error assumes spectral fits to the velocity of an ion are good to within 500 km s$^{-1}$. The black horizontal line represents the average value of data points (excluding SN 2001V, 2006bt and 2008ha) and the dashed horizontal lines indicate a 10\% difference from the mean.}
\label{fig:ratio}
\end{figure}

In Figure~\ref{fig:ratio}, we have the ratio of Doppler velocities,
$v$(\ion{C}{2} $\lambda$6580)/$v$(\ion{Si}{2} $\lambda$6355), derived from our
\texttt{SYNOW} fits plotted versus days relative to maximum light. This shows
that (1) for an individual SN, the ratio remains at a fairly sustained value
over time and (2) the different velocity ratios among the SNe lie roughly
within the same region and are similar to within $\pm$10\%. We have ignored the
three outliers because one is in a region of the plot where line blending
obscures the supposed \ion{C}{2} $\lambda$6580 detection (SN 2001V; see
\S6.2.2) and the other two objects suggest that the carbon is clumpy and not
along the line-of-sight of the observer (SN 2006bt and SN 2008ha). 

The notion of an optically thick photosphere in SNe is generally a good
assumption,  even though the line forming region may extend 500 km s$^{-1}$ in
either direction. Therefore, for a given spectrum, any velocities of an ion
that are measured to be fairly below that of the photospheric velocity,
v$_{phot}$, may indicate ejecta asymmetries. That is, any observed discrepancy
between v$_{phot}$ and v$_{C II}$ could be explained if the actual velocity of
\ion{C}{2} is the same as v$_{phot}$ but instead of forming in a shell at the
observed velocity, the carbon is in a clump at v$_{phot}$ and offset by an
angle, $\theta$, from the line of sight. In particular, we can estimate this
projection angle if v$_{C II}$ $<$ v$_{phot}$. For SN 2006bt and SN 2008ha we
calculate this projection angle to be $\sim$50$^{\circ}$ and 60$^{\circ}$,
respectively, where our result for SN 2006bt is in agreement with \cite{F10b}. 

If the dominant \ion{C}{2} behavior is due to asymmetrical
distributions of a single localized clump of unburned material, one could
expect there to be more scatter below the mean that is presented in
Figure~\ref{fig:ratio} since an arbitrary orientation of the clump relative to
the observer ought to lead to more cases where $v$(\ion{C}{2}
$\lambda$6580)/$v$(\ion{Si}{2} $\lambda$6355) $<$ 1 (like SN 2006bt). Instead,
what we find for the SNe in our sample is that the distribution of carbon-rich
material is consistent with a layered or hemispheric geometry. 

Another surprising aspect of Figure~\ref{fig:ratio} is that the candidate Super-Chandra
SNe Ia reside in the same region of the plot as the other objects.
\cite{Scalzo10} suggested that the large \ion{C}{2} $\lambda$6580 feature,
concurrent with low velocities, could be explained by invoking a pre-explosion
envelope of progenitor material originating from the merger of two white
dwarfs. In this scenario, the explosion is inhibited by and loses kinetic
energy to the envelope, ionizing the shell of surrounding carbon. Whatever the
nature of carbon-rich regions in these three SNe Ia, it is also constrained
by the value of $v$(\ion{C}{2} $\lambda$6580)/$v$(\ion{Si}{2} $\lambda$6355). 
 
\subsubsection{Consequences of Line Blending}

Toward the uppermost region of Figure~\ref{fig:ratio}, there is a noticeable
lack of highly detached \ion{C}{2} objects with time-series coverage that
exhibit pre-maximum \ion{C}{2} $\lambda$6580 absorption features. This does not
necessarily imply that carbon-rich regions in SNe Ia are absent above a
particular velocity. Rather, the ``missing'' subset of objects may be a result
of SNe Ia with lower density carbon-rich regions further out and/or radiative
transfer selection effects; e.g. line blending. 

Most of the SNe Ia in our sample that exhibit \ion{C}{2} $\lambda$6580
absorption features are of the LVG subtype while only one is an HVG event (SN
2009ig). This is either a real trend of \ion{C}{2} $\lambda$6580 absorption
features and therefore a diagnostic of SN Ia diversity, or a result of line
blending due to very high velocities of carbon-rich regions. In regards to the
latter cause, some SNe Ia might contain a \ion{C}{2} $\lambda$6580 absorption
feature in their spectra but its presence is obscured by the \ion{Si}{2}
$\lambda$6355 absorption trough, particularly for HVGs. 

Assuming that the scatter of $v$(\ion{C}{2} $\lambda$6580)/$v$(\ion{Si}{2}
$\lambda$6355) values in Figure~\ref{fig:ratio} is the same for both HVG and
LVG subtypes, and assuming that HVGs span a $v_{phot}$-space from 14,000 to
16,000 km s$^{-1}$, then the minimum of any \ion{C}{2} $\lambda$6580 signature
would be between 6120$-$6280 \AA. If weak and blue-shifted to these
wavelengths, the \ion{C}{2} $\lambda$6580 absorption would most likely go
undetected. If the \ion{C}{2} optical depth were large enough, then the most
direct evidence for hidden \ion{C}{2} $\lambda$6580 would be whether or not an
associated absorption from \ion{C}{2} $\lambda$7234 could be seen between
6730$-$6900 \AA.  

This raises the question: At what velocity will \ion{C}{2} $\lambda$6580
completely blend with \ion{Si}{2} $\lambda6580$? In
Figure~\ref{fig:samplefits}, we presented observed \ion{C}{2} $\lambda$6580
blending scenarios where the signature is weak and nearly hidden, such as
possibly observed in SN 2001V and 2009ig. These objects suggest that \ion{C}{2}
$\lambda$6580 absorption features can in some cases be obscured via line
blending. Such a ``hidden'' \ion{C}{2} signature would be seen in the form of
an absorption on the shoulder on the blue wing of the \ion{Si}{2} $\lambda$6355
emission component. 

\begin{figure}
\centering
\includegraphics[width=12.8cm]{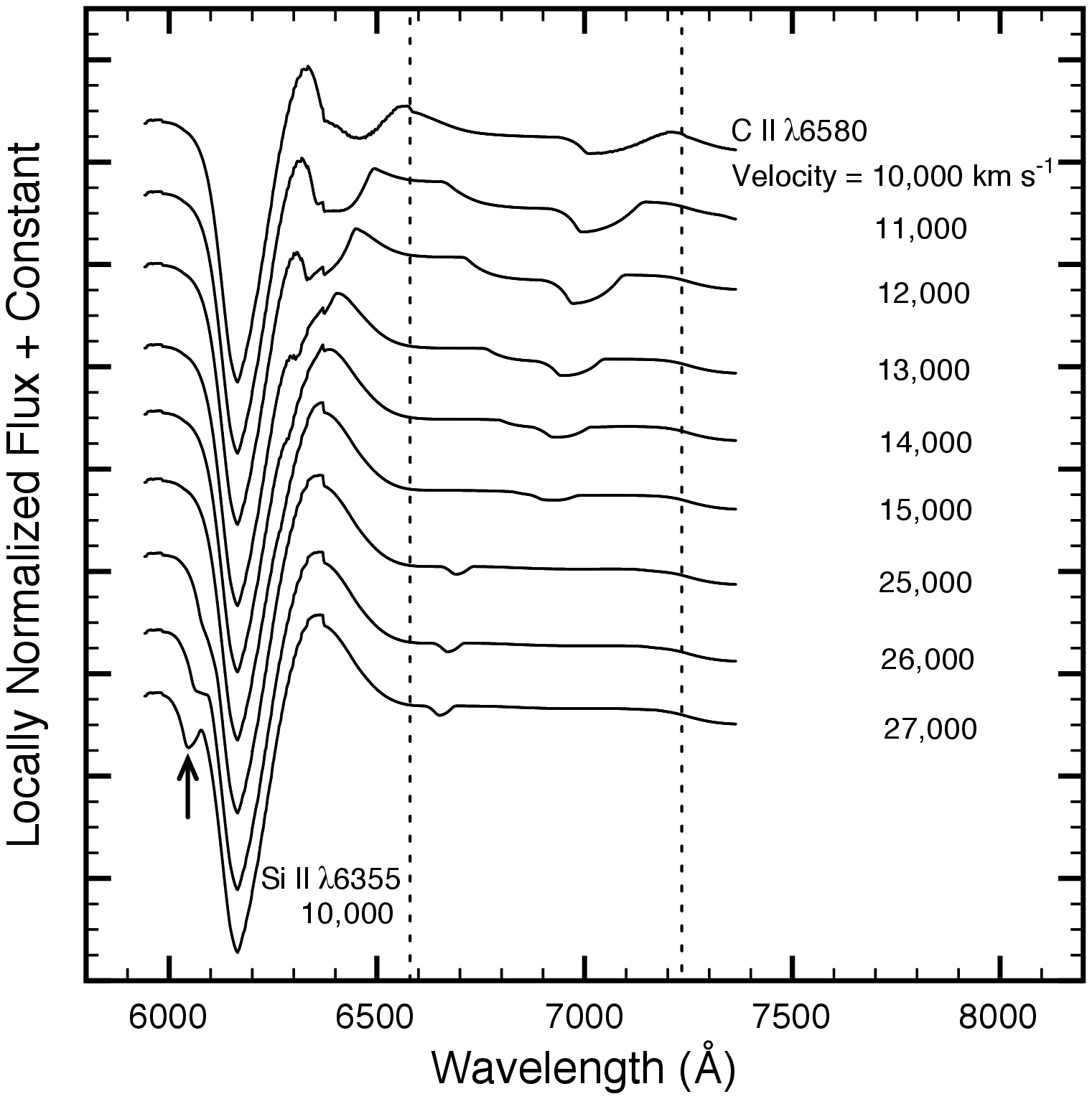}
\centering
\includegraphics[width=12.8cm]{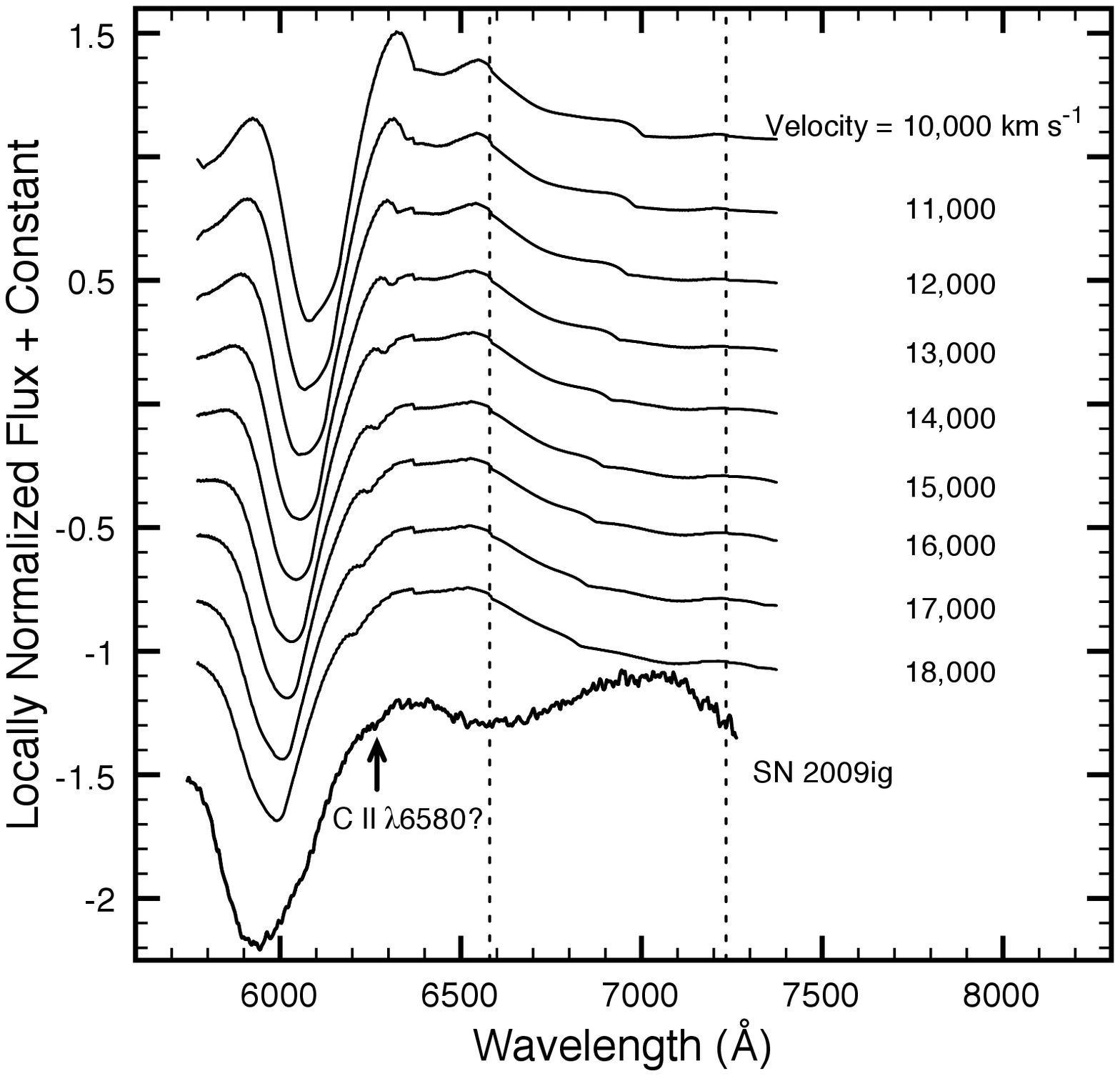}
\caption{(top) A series of synthetic spectra show an evolution of blending scenarios between \ion{C}{2} $\lambda$6580 and \ion{Si}{2} $\lambda$6355 as the velocity of \ion{C}{2} is increased from 10,000 to 15,000 km s$^{-1}$. The two spectra at the bottom represent a case where the \ion{C}{2} is present at high velocities. Note that even if the 6580-line is obscured, the 7234-line is blueshifted to the rest frame position of the 6580-line. Three high-velocity scenarios are also shown to note the velocity at which \ion{C}{2} $\lambda$6580 emerges from the blue wing of \ion{Si}{2} $\lambda$6355. (bottom) Like the top figure, but instead \ion{C}{2} and \ion{Si}{2} are at the same velocity and increased in sequence together while holding $\tau$ fixed. The \ion{Si}{2} profiles are made from two-components of Si separated by 4000 km s$^{-1}$. The day $-$14 spectrum of SN 2009ig is shown for comparison.}
\label{fig:sic}
\end{figure}

A series of synthetic spectra that include only blends of \ion{Si}{2} $\lambda$6355 and \ion{C}{2} $\lambda$6580 line profiles are shown in Figure~\ref{fig:sic} (top panel). For these \texttt{SYNOW} spectra, the velocity of \ion{Si}{2} was fixed at 10,000 km s$^{-1}$ while the velocity of \ion{C}{2} was increased from this 10,000 km s$^{-1}$ by 1000 km s$^{-1}$ increments up to a velocity of 15,000 km s$^{-1}$. We simultaneously decreased the optical depth in keeping consistency with the observed profile shapes. At a velocity of 15,000 km s$^{-1}$, \ion{C}{2} $\lambda$6580 loses its discernibility as a feature and remains hidden until \ion{C}{2} reaches a velocity of 27,000 km s$^{-1}$. At present there are no observations reporting a \ion{C}{2} $\lambda$6580 feature appearing blue-ward of the \ion{Si}{2} $\lambda$6355 absorption. 
 
Aside from being able to account for the observed variety of \ion{C}{2} absorption profile shapes and velocities with respect to \ion{Si}{2} $\lambda$6355, in Figure~\ref{fig:sic} we also show that the shoulder effect can be reproduced when \ion{C}{2} is detached with a velocity $\sim$4000 km s$^{-1}$ greater than that of \ion{Si}{2}. This is consistent with our fit for SN 2001V where we have \ion{C}{2} and \ion{Si}{2} at 12,000 and 8000 km s$^{-1}$, respectively. 

In Figure~\ref{fig:sic} (bottom panel), we show a series of synthetic spectra where \ion{C}{2} and \ion{Si}{2} are placed at the same velocity and increased in sequence from 10,000 to 18,000 km s$^{-1}$. We modeled the \ion{Si}{2} profiles using the two-component method so as to reproduce HVG-like properties. As can be seen, the \ion{C}{2} $\lambda$6580 absorption feature can be easily obscured at higher velocities. The proposed weak \ion{C}{2} $\lambda$6580 feature for SN 2009ig lies atop the emission component of \ion{Si}{2} $\lambda$6355 and is consistent with this series of synthetic spectra. 

It is generally thought that the infrequent number of HVGs where \ion{C}{2} $\lambda$6580 is detected suggests an environment of sparse carbon-rich regions \citep{Pignata08}. However, if the proposed \ion{C}{2} $\lambda$6580 signature in SN 2009ig is correct, then our fits suggest that the feature could be weak based on line blending, and not necessarily due to a low carbon abundance (or lower than that of LVGs) alone. 

\subsection{Asymmetries in the Distribution of \ion{C}{2} Regions}

Because the ejected material follow a homologous expansion law ($v \propto r$),
the degree to which electron scattering will polarize emergent light across a
spectral line can provide geometrical information on the material between the
observer and the photosphere \citep{SS82,WW08}. Spectropolarimetry of some SNe
Ia have revealed that the distribution of the outermost ejected material has an
overall deviation from spherical symmetry of up to $\sim$10\%
\citep{Hoflich91,Wang97}. Specifically, optical \ion{Si}{2} P-Cygni profiles
are observed to have peak polarization values of 0.3$-$2.0\%, five days before
maximum light \citep{Wang07}. Other absorption features, such as the
\ion{Ca}{2} IR triplet, have also been observed to be strongly polarized
\citep{Wang03,Kasen03}. 

Of the 19 SNe Ia in our sample, only SN 1994D, 1999by and 2005hk have nearly
simultaneous spectroscopic and polarization data during the presence of a
\ion{C}{2} $\lambda$6580 absorption feature. Polarization levels for these
objects across \ion{Si}{2} $\lambda$6355 were reported as insignificant
($\sim$0.3\% for SN 1994D, \citealt{Wang96}; $\sim$0.4\% for SN 1999by,
\citealt{Howell01}; $\sim$0.4\% for SN 2005hk, \citealt{Chornock06}), whereas
the degree of polarization was even less in the region of the \ion{C}{2}
$\lambda$6580 signature. 

Regarding SN 1999by, the three spectropolarimetric observations of \cite{Howell01} were combined
from data taken on days $-$2, $-$1 and 0 in order to increase the S/N ratio
(see their Fig.\ 2). The resultant spectrum exhibited the same
6300 \AA\ absorption feature that we attributed to \ion{C}{2} $\lambda$6580 in
the day $-$5, $-$4 and $-$3 spectra of \cite{Garnavich04}, but the concurrent
polarimetry data of \cite{Howell01} did not show any significant amount of
polarization near the feature. This would suggest that the \ion{C}{2} regions
are roughly spherical in at least these objects. However, because
spectropolarimetric observations are most useful when a spectral line is strong
and the S/N is high, it is difficult to use the weak
\ion{C}{2} $\lambda$6580 absorption features of SNe Ia to infer asymmetries of
carbon-rich regions with polarization data. Despite this, possible asymmetries
of carbon-rich regions might be gleaned through other means. 

Recently, \cite{Maedanature} discussed SN Ia diversity in the context of global asymmetries of ejected material. They argued that the HVG and LVG subtypes constitute a picture of SNe Ia in terms of an off-center delayed-detonation at different viewing angles. In this scenario, an off-center ignition is followed by the propagation of a sub-sonic deflagration flame that imprints an off-set distribution of high density ash. Once the transition to detonation occurs, a super-sonic flame only successfully burns material in the lower density regions and is effectively screened by some of the deflagration ash. What remains is a lopsided distribution of burning products, resulting in a hemispheric asymmetry of the ejecta. Similarly, \cite{Maund10} reached the same conclusion regarding SN Ia diversity after examining a possible relationship between HVG/LVG subtypes and the possible asymmetrical distribution of photospheric \ion{Si}{2}.

In this model, LVGs correspond to viewing the hemisphere of ejecta that coincides with the off-center ignition, whereas HVGs are the contrary with an opening angle of $\sim$70$-$75$^{\circ}$. Since we do not see a predominate number of HVGs with conspicuous \ion{C}{2} $\lambda$6580 features in our sample, this is consistent with the above model, owing to the fact that detonation waves ought to leave little unburned material behind on the HVG side. If \cite{Maedanature} is correct, and if our sample of SN Ia that exhibit \ion{C}{2} $\lambda$6580 is representative, then this suggests that the filling factor of carbon in the LVG hemisphere is less than unity. 

\section{Conclusions}

In an effort to better understand the frequency and general properties of
\ion{C}{2} absorption features, we examined \ion{C}{2} $\lambda$6580 signatures
in the pre-maximum spectra of 19 SNe Ia, of which included 14 ``normal'' SNe Ia, three
possible Super-Chandra SNe Ia, and two 2002cx-like events. Using \texttt{SYNOW} to
produce synthetic spectra, we modeled observed $\sim$6300 \AA\ absorption
features as a \ion{C}{2} $\lambda$6580 P-Cygni profile blended with that of
\ion{Si}{2} $\lambda$6355. Through our \texttt{SYNOW} model fits we estimated the
\ion{C}{2} expansion velocities for a variety of objects. Below is a summary of
our major findings:

\begin{enumerate}

\item A survey of the optical spectra of 68 objects published in the literature
since 1983 January 1 indicates that up to $\sim$30\% of SNe Ia display an absorption
feature near 6300 \AA\ that can be attributed to \ion{C}{2} $\lambda$6580.
While this percentage is likely biased, it does suggest that \ion{C}{2}
$\lambda$6580 absorption features are more common than was previously suspected
\citep{Thomas07}. If spectroscopic observations of SNe Ia are obtained more
than $\sim$1 week before maximum light, we suspect an even larger fraction of
SNe Ia of all subtypes may show identifiable \ion{C}{2} absorption signatures. 

\item A greater frequency of \ion{C}{2} $\lambda$6580 absorption features appear in the low-velocity gradient subtypes (LVGs) compared to
high-velocity gradient events (HVGs). This is in line with the interpretation
of \cite{Maedanature}, supporting the idea that part of SN Ia diversity can be
accounted for by viewing angle and off-center ignition effects.

\item The influence of the temperature of the carbon-rich region on the
incidence of \ion{C}{2} $\lambda$6580 signatures is most evident for the 14
``normal'' SNe Ia in our survey. The frequency of \ion{C}{2} $\lambda$6580
signatures peaks with the Core-Normal (CN) SNe Ia, while the Cool (CL) and
Shallow-Silicon (SS) subtypes exhibit fewer \ion{C}{2}
$\lambda$6580 detections. This result is consistent with the effective
temperature sequence of \cite{Nugent95}.

\item We find the values of $v$(\ion{C}{2} $\lambda$6580)/$v$(\ion{Si}{2}
$\lambda$6355) among 16 of the SNe Ia in our \ion{C}{2} sample are 
similar to within $\pm$10\%.  Assuming the minima of \ion{Si}{2} $\lambda$6355
absorption features are an appropriate measure of photospheric velocities prior
to maximum light, then the small number of cases where $v$(\ion{C}{2}
$\lambda$6580)/$v$(\ion{Si}{2} $\lambda$6355 $<$ 1 could be indicative of
either a layered distribution or multiple clumps with a comparable filling
factor.

\end{enumerate}

One of the most interesting results of this study is that \ion{C}{2}
$\lambda$6580 absorption features might be in many, if not most, early-epoch SN
Ia spectra. We initially set out to investigate what we suspected was a small
number of known detections of \ion{C}{2} $\lambda$6580. However, when we
examined the 6300 \AA\ region in a relatively large sample of objects, we
discovered that weak signatures of \ion{C}{2} $\lambda$6580 were often
detectable and could influence the observed \ion{Si}{2} $\lambda$6355 spectral
profile. 

One way to test for a high frequency of \ion{C}{2} $\lambda$6580 features in
pre-maximum SNe Ia spectra would be to obtain high S/N observations of the
region between 6700$-$7200 \AA\ in order to look for the relatively weaker
\ion{C}{2} $\lambda$7234 line that should also be present. In addition, a data
set of \ion{C}{2} $\lambda$6580 observations with wavelength coverage that
encompasses the \ion{O}{1} $\lambda$7774 absorption feature would allow for
investigating any correlations between C and O spectral line properties. 

Understanding the presence of \ion{C}{2} in SN Ia spectra may prove to be a
valuable diagnostic of SN Ia diversity and therefore a possible means by which
to probe the underlying explosion mechanism. Utilizing the absolute strength of
the \ion{C}{2} $\lambda$6580 line in order to extract abundance information
will require comparison to synthetic spectra calculations based on
hydrodynamical modeling. A  detailed comparative study that can extract C
abundance information, derived from \ion{C}{2} $\lambda$6580 absorption
features, promises to produce a wealth of insight regarding carbon-rich
regions, and therefore constrain the parameter space of hydrodynamical
modeling. 

\acknowledgements

We are grateful to Laura Kay for obtaining some of the spectra of SN 2010Y
and thank Eddie Baron, David Branch, and Andy Howell for helpful comments on an earlier
draft of this paper. J.V. has received support from NSF Grant AST-0707669,
Texas Advanced Research Program grant ASTRO-ARP-0094
and Hungarian OTKA Grant K76816.


\clearpage
\newpage


\clearpage
\newpage

\clearpage
\newpage

\clearpage
\newpage

\clearpage
\newpage

\clearpage
\newpage

\clearpage
\newpage

\clearpage
\newpage

\clearpage
\newpage

\clearpage
\newpage

\begin{deluxetable}{ccclcc}
\tablenum{1}
\tablecolumns{6}
\tabletypesize{\small}
\setlength{\tabcolsep}{0.15in}
\tablewidth{0in}
\tablecaption{Summary of Observations}
\tablehead{
\colhead{Supernova}       & 
\colhead{Host Galaxy}     & 
\colhead{Redshift}        & 
\colhead{Observation}     & 
\colhead{Epoch}           & 
\colhead{Telescope}       \\
\colhead{Name}            & 
\colhead{}                & 
\colhead{(km s$^{-1}$)}   & 
\colhead{Date}            & 
\colhead{(days)}          & 
\colhead{Instrument}
}

\startdata
SN 2010Y & NGC 3392 & 3256 & 2010 Feb 09 & -7 & MDM/CCDS \\
\omit & \omit & \omit & 2010 Feb 10 & -6 & MDM/CCDS \\
\omit & \omit & \omit & 2010 Feb 13 & -3 & MDM/CCDS \\
\omit & \omit & \omit & 2010 Feb 14 & -2 & HET/LRS \\
\omit & \omit & \omit & 2010 Feb 17 & +1 & HET/LRS \\
SN 2010ai & SDSS J125925.04+275948.2 & 5507 & 2010 Mar 11 & -10 & HET/LRS \\
\omit & \omit & \omit & 2010 Mar 13 & -8 & HET/LRS \\
PTF10icb & \nodata & \nodata & 2010 June 3 & -10 & HET/LRS \\                
\enddata
\end{deluxetable}

\clearpage
\newpage

\begin{deluxetable}{lccccclcccc}
\tablenum{2}
\tabletypesize{\small}
\setlength{\tabcolsep}{0.07in}
\tablecolumns{11}
\tablewidth{0in}
\tablecaption{SN Ia Sample}
\tablehead{
\colhead{SN} &
\colhead{Sub-type\tablenotemark{a}} &
\colhead{\ion{C}{2}?} &
\colhead{Epoch} &
\colhead{Spectrum} &
\colhead{} &
\colhead{SN} &
\colhead{Sub-type} &
\colhead{\ion{C}{2}?} &
\colhead{Epoch} &
\colhead{Spectrum} \\
\colhead{Name} &
\colhead{} &
\colhead{} &
\colhead{(days)} &
\colhead{Source\tablenotemark{e}} &
\colhead{} &
\colhead{Name} &
\colhead{} &
\colhead{} &
\colhead{(days)} &
\colhead{Source} 
}

\startdata
\cutinhead{LVG}
SN 1990N\tablenotemark{b} & CN & Definite & -14 & M01 & & SN 1999dq & SS & No & -9.5 & M08 \\
SN 1991T\tablenotemark{b} & SS & Possible & -13 & F99 & & SN 1999ee & SS & Uncertain & -9 & Mazz05 \\
SN 1994D\tablenotemark{b,}\tablenotemark{c} &  CN & Probable & -11 & P96 & & SN 1999gp & SS & No & -4.5 & M08 \\
SN 1995D & CN & Uncertain & +0 & S96 & & SN 2000E & SS & Possible & -6 & V03 \\
SN 1996X\tablenotemark{b,}\tablenotemark{c} & CN & Probable & -4 & S01 & & SN 2001V\tablenotemark{b} & SS & Possible & -7 & M08 \\
SN 1997br\tablenotemark{b} & SS & No & -9 & L99 & & SN 2001el & CN & Probable & -9 & Matt05 \\
SN 1997dt & CN & Uncertain & -10 & M08 & & SN 2003cg & CN & Possible & -8.5 & E-R06 \\
SN 1998V & CN & Uncertain & +0.5 & M08 & & SN 2003du\tablenotemark{b,}\tablenotemark{c} & CN & Probable & -11 & S07 \\ 
SN 1998ab & SS & No & -7.5 & M08 & & SN 2005cf & CN & Uncertain & -12 & W09 \\
SN 1998aq\tablenotemark{b} & CN & Probable & -9 & B03 & & SN 2005cg & SS & No & -10 & Q06 \\
SN 1998bu & CN & Probable & -7 & H00 & & SN 2005hj & SS & No & -6 & Q07 \\
SN 1998es & SS & No & -10 & M08 & & SN 2006D & CN & Definite & -7 & T07\\
SN 1999aa & SS & No & -11 & Gara04 & & SN 2010ai & CN & Probable & -10 & this work \\
SN 1999ac\tablenotemark{b} & SS & Possible & -15 & Gara05 & & PTF10icb & CN & Probable & -10 & this work \\
\cutinhead{HVG}
SN 1981B & BL & No & 0 & B83 & & SN 2000fa & BL & Uncertain & -10 & M08 \\
SN 1984A & BL & No & -7 & B89 & & SN 2002bo & BL & No & -13 & S05 \\
SN 1992A & BL & Probable & -6.5 & K93 & & SN 2002dj & BL & No & -11 & P08 \\
SN 1997do & BL & No & -11 & M08 & & SN 2002er & BL & Possible & -11 & K05 \\
SN 1998dh & BL & No & -9 & M08 & & SN 2004dt\tablenotemark{b} & BL & No & -7& W06 \\
SN 1998ec & BL & No & -2.5 & M08  & & SN 2006X & BL & Possible & -7 & Y09 \\
SN 1999cc & BL & No & -3 & M08 & & SN 2007gi & BL & Possible & -7.5 & Z10 \\
SN 1999cl & BL & No & -7.5 & M08 & & SN 2009ig & BL & Possible & -14 & M11, in prep. \\
SN 1999ej & BL & No & -0.5 & M08 & & & & \\
\cutinhead{FAINT}
SN 1986G & CL & Possible & -6 & P87 & & SN 2000cn & CL & Uncertain & -9 & M08 \\
SN 1989B & CL & Possible & -7 & W94 & & SN 2000dk & CL & Possible & -4.5 & M08 \\
SN 1991bg & CL & No & +1 & T96 & & SN 2004eo & CL & No & -11 & P07 \\
SN 1997cn & CL & No & 0 & T98 & & SN 2005bl\tablenotemark{b} & CL& Probable & -6 & T08 \\
SN 1998bp & CL & No & -2.5 & M08 & & SN 2006bt & CL & Probable & -3 & F10b \\
SN 1998de & CL & No & -6.5 & M08 & & SN 2010Y & CL & Definite & -7 & this work \\
SN 1999by\tablenotemark{b,}\tablenotemark{c} & CL & Possible & -5 & Garn04 & & & & & \\
\cutinhead{SC\tablenotemark{d}}
SN 2003fg & \nodata & Definite & & H06 & & SN 2006gz\tablenotemark{b} & \nodata & Definite & -14 & H07 \\
SN 2007if & \nodata & Definite & -9 & S10 & & SN 2009dc & \nodata & Definite & -8 & unpublished \\
\cutinhead{misc.}
SN 2000cx & \nodata & Possible & -3 & L01 & & SN 2005hk\tablenotemark{b,}\tablenotemark{c} & \nodata & Possible & -5 & S08 \\
SN 2002bj & \nodata & Definite & +7 & P10 & & SN 2007qd & \nodata & Possible & +3 & M10 \\
SN 2002cx  & \nodata & Probable & -4 & L03 & & SN 2008ha & \nodata & Definite & -1 & F09, F10a \\
\enddata
\tablenotetext{a}{SN type notation of \cite{Branch06}, where CN = `core normal', SS = `shallow silicon', CL = `cool', and BL = `broad line'. The similar subtypes of \cite{Benetti05} have been used to separate to these SNe Ia into subclasses.}
\tablenotetext{b}{Taken from The Supernova Spectrum Archive, SuSpect.}
\tablenotetext{c}{Objects with both published spectropolarimetry data and \ion{C}{2} $\lambda$6580 absorption signatures: 1994D; \cite{Wang96}, 1996X; \cite{Wang97}, 1999by; \cite{Howell01}, 2003du; \cite{Leonard05}, 2005hk; \cite{Chornock06}.}
\tablenotetext{d}{SC = Super-Chandra Candidates}
\tablenotetext{e}{References$-$(B89) \citealt{Barbon89}; (B83) \citealt{Branch83}; (B03) \citealt{Branch03}; (E-R06) \citealt{Elias-Rosa06}; (F99) \citealt{Fisher99}; (F09) \citealt{F09}; (F10a) \citealt{F10a}; (F10b) \citealt{F10b}; (Gara04) \citealt{Garavini04}; (Gara05) \citealt{Garavini05}; (Garn04) \citealt{Garnavich04}; (H00) \citealt{Hernandez00}; (H07) \citealt{Hicken07}; (H06) \citealt{Howell06}; (K93) \citealt{Kirshner93}; (K05) \citealt{Kotak05}; (L99) \citealt{Li99}; (L01) \citealt{Li01}; (L03) \citealt{Li03}; (M11) \citealt{Marion11}; (M08) \citealt{Matheson08}; (Matt05) \citealt{Mattila05}; (M01) \citealt{Mazzali01}; (Mazz05) \citealt{Mazzali05}; (M10) \citealt{McClelland10}; (M11) Marion et al. 2011; (P07) \citealt{Pastorello07}; (P96) \citealt{Patat96}; (P87) \citealt{Phillips87}; (P08) \citealt{Pignata08}; (P10) \citealt{Poznanski10}; (Q06) \citealt{Quimby06}; (Q07) \citealt{Quimby07}; (S08) \citealt{Sahu08}; (S01) \citealt{Salvo01}; (S96) \citealt{Sadakane96}; (S10) \citealt{Scalzo10}; (S07) \citealt{Stanishev07}; (S05) \citealt{Stehle05}; (T08) \citealt{Taubenberger08}; (T07) \citealt{Thomas07}; (T96) \citealt{Turatto96}; (T98) \citealt{Turatto98}; (V03) \citealt{Valentini03}; (W06) \citealt{Wang06}; (W09) \citealt{Wang09}; (W94) \citealt{Wells94}; (Y09) \citealt{Yamanaka09a}; (Z10) \citealt{Zhang10}.}
\end{deluxetable}

\clearpage
\newpage

\begin{deluxetable}{lrrrrrrrr}
\tablenum{3}
\tablecolumns{7} 
\tablewidth{0pt} 
\tablecaption{\texttt{SYNOW} Fit Parameters}
\tablehead{\colhead{Supernova}                                 &
           \colhead{Epoch}                                     &
           \colhead{$v_{phot}$}                                &
           \multicolumn{2}{c}{\ion{C}{2}\tablenotemark{a}}     &
           \multicolumn{2}{c}{\ion{Si}{2}\tablenotemark{a}}    \\
           \colhead{Name}                                      &        
           \colhead{(days)}                                    &
           \colhead{(km s$^{-1}$)}                             &
           \colhead{Velocity}                                  &               
           \colhead{$\tau$}                                    &            
           \colhead{Velocity}                                  & 
           \colhead{$\tau$}                                    & 
           \colhead{}                                          & 
           \colhead{}                                          }
\startdata
SN 1990N & -14 & 13,000 & 16,000 & 0.60 & $>$14,000\tablenotemark{b} & 0.50 \\
                    & -8 & 11,000 & 14,000 & 0.25 & 11,000 & 1.40 \\
                    & -2 & 11,000 & 14,000 & 0.15 & 11,000 & 16.0 \\
                    & 2 & 11,000 & 14,000 & 0.35 & 11,000 & 16.0 \\ 
SN 1994D & -11 & 14,000 & 14,000 & 1.30 &  $>$14,000 & 14.0 \\
                    & -10 & 13,000 & 14,000 & 0.45 &  $>$13,000 & 12.0 \\
                    & -8 & 13,000 & 14,000 & 0.45 &  $>$13,000 & 6.00 \\
                    & -5 & 12,000 & 14,000 & 0.20 &  $>$12,000 & 6.00 \\
SN 1996X & -4 & 13,500 & 13,500 & 0.60 & 13,500 & 15.0 \\
                    & -2 & 13,000 & 13,000 & 0.40 & 13,000 & 13.0 \\
                    & -1 & 12,000 & 13,000 & 0.04 & 13,000 & 8.00 \\
SN 1998aq & -9 & 12,000 & 15,000 & 0.20 & 13,000 & 2.10  \\
                      & -8 & 12,000 & 14,000 & 0.20 & 13,000 & 2.10  \\
                      & 0 & 12,000 & 13,500 & 0.60 & 12,000 & 9.00  \\
SN 1999ac & -15 & 13,000 & 17,000 & 0.40  & 15,000 & 2.00 \\
                     & -9   & 14,000 & 17,500 & 0.30 & 14,000 & 3.00 \\  
SN 1999by & -5   & 12,000 & 14,500 & 0.50 & 12,000 & 12.0 \\
                      & -4   & 12,000 & 14,500 & 0.30 & 12,000 & 13.0 \\
                      & -3   & 12,000 & 14,500 & 0.30 & 12,000 & 14.0 \\
SN 2001V & -7 & 8000 & 12,000 & 0.15 & 8000 & 1.60  \\
                    & -6 & 8000 & 12,000 & 0.15 & 8000 & 1.60  \\
                    & -4 & 8000 & 12,500 & 0.15 & 8000 & 1.80  \\
SN 2003du & -13 & 14,000 & 14,000 & 0.60 & $>$14,000 & $>$4.00 \\
                      & -11 & 12,000 & 14,000 & 0.60 & $>$12,000 & $>$4.50 \\
SN 2005hk & -5 & 6000 & 7500 & 0.07 & 7000 & 0.30 \\ 
                     & 0 & 6000 & 7500 & 0.14 & 7000 & 0.70  \\
SN 2006D & -7 & 11,000 & 14,500 & 0.30 & 13,000 & 0.70 \\
                    & -5 & 11,000 & 14,500 & 0.20 & 13,000 & 0.70 \\
SN 2006bt & -4 & 6500 & 6500 & 0.30 & $>$11,000 & $>$0.50 \\
                    & -3 & 6500 & 6500 & 0.25 & $>$11,000 & $>$0.70 \\
SN 2006gz & -14 & 12,000 & 18,500 & 0.50 & 16,500 & 1.10 \\
                     & -12 & 12,000 & 18,500 & 0.40 & 16,500 & 1.50 \\
                     & -10 & 12,000 & 17,500 & 0.30 & 16,000 & 1.60 \\
SN 2007if & -9 & 8500 & 9500 & 0.25 & 8500 & 0.30 \\
SN 2008ha\tablenotemark{c} & -1 & 1000 & 1000 & 1.00 & 2500 & 10.0 \\
SN 2009dc & -8 & 8000 & 11,000 & 0.50 & 10,000 & 0.80 \\
                     & -7 & 8000 & 10,000 & 0.30 & 9500 & 0.50 \\
                     & -6 & 8000 & 10,000 & 0.30 & 9500 & 0.50 \\
                     & -5 & 7000 & 9500 & 0.30 & 9000 & 0.50 \\
                     & -3 & 7000 & 9500 & 0.30 & 8500 & 0.50 \\
                     & -2 & 6500 & 9000 & 0.30 & 8000 & 0.50 \\
                     & -1 & 6500 & 9000 & 0.30 & 8000 & 0.50 \\
                     & 0 & 6500 & 8500 & 0.40 & 8500 & 1.10 \\
                     & 2 & 6500 & 8500 & 0.30 & 8000 & 0.60 \\
                     & 3 & 6000 & 7500 & 0.30 & 7500 & 0.70 \\
                     & 4 & 6000 & 7500 & 0.30 & 7500 & 0.70 \\
                     & 5 & 6000 & 7500 & 0.30 & 7500 & 0.70 \\
                     & 6 & 5500 & 7500 & 0.30 & 7000 & 0.70 \\
                     & 7 & 5500 & 7500 & 0.30 & 6500 & 0.70 \\
                     & 8 & 6000 & 7500 & 0.30 & 7000 & 0.70 \\
                     & 9 & 5000 & 6000 & 0.20 & 7000 & 1.00 \\
                     & 15 & 5000 & 5000 & 0.20 & 6500 & 0.80 \\
SN 2009ig & -14 & 16,000 & 16,000 & 1.00 & $>$16,000 & $>$24  \\
                     & -13 & 16,000 & 16,000 & 0.50 & $>$16,000 & $>$24  \\
SN 2010Y & -7 & 12,000 & 15,000 & 0.50 & 14,000 & 6.00  \\
                    & -6 & 12,000 & 15,000 & 0.50 & 14,000 & 5.00  \\
                    & -3 & 12,000 & 14,500 & 0.50 & 13,000 & 10.0  \\
                    & -2 & 11,500 & 14,000 & 0.50 & 12,500 & 18.0 \\
                    & 1 & 11,000 & 14,000 & 0.20 & 12,000 & 20.0  \\
SN 2010ai & -10 & 13,500 & 14,000 & 1.00 & 13,500 & 2.00  \\
		  & -8 & 12,500 & 14,000 & 0.50 & 12,500 & 1.50 \\
PTF10icb & -10 & 11,000 & 14,500 & 0.10 & $>$12,000 & $>$1.10  \\
\enddata
\tablenotetext{a}{Fits made with T$_{exc}$=10,000 K and v$_{e}$=1000 km s$^{-1}$.}
\tablenotetext{b}{Used two-components of \ion{Si}{2} to fit broad \ion{Si}{2} $\lambda$6355 P-Cygni profiles.}
\tablenotetext{c}{See \cite{Valenti09} for an alternate \texttt{SYNOW} fit where the origin of explosion is interpreted to be a core-collapse supernova.}
\end{deluxetable}

\end{document}